\documentstyle[pre,aps,eqsecnum,epsf,rotate,twocolumn]{revtex}
\begin{document}
\draft
\tighten
\title
{\LARGE Two--Loop Renormalization Group Analysis 
of the \\ Burgers -- Kardar-Parisi-Zhang Equation}
\author{Erwin Frey}
\address{
Institut f\"ur Theoretische Physik,
Physik-Department der Technischen Universit\"at M\"unchen, \\
James-Franck-Stra\ss e, D-85747 Garching, Germany}
\author{Uwe Claus T\"auber}
\address{
Lyman Laboratory of Physics,
Harvard University,
Cambridge, Massachusetts 02138, U.S.A.}
\date{\today}
\maketitle
\widetext

\begin{abstract}
A systematic analysis of the Burgers---Kardar--Parisi--Zhang equation in $d+1$
dimensions by dynamic renormalization group theory is described. The fixed 
points and exponents are calculated to two--loop order. We use the dimensional
regularization scheme, carefully keeping the full $d$ dependence originating
from the angular parts of the loop integrals. For dimensions less than $d_c=2$ 
we find a strong--coupling fixed point, which diverges at $d=2$, indicating 
that there is non--perturbative strong--coupling behavior for all $d \geq 2$. 
At $d=1$ our method yields the identical fixed point as in the one--loop 
approximation, and the two--loop contributions to the scaling functions are 
non--singular. For $d>2$ dimensions, there is no finite strong--coupling fixed 
point. In the framework of a $2+\epsilon$ expansion, we find the dynamic 
exponent corresponding to the unstable fixed point, which describes the 
non--equilibrium roughening transition, to be $z = 2 + {\cal O} (\epsilon^3)$,
in agreement with a recent scaling argument by Doty and Kosterlitz. Similarly,
our result for the correlation length exponent at the transition is $1/\nu = 
\epsilon + {\cal O} (\epsilon^3)$. For the smooth phase, some aspects of the 
crossover from Gaussian to critical behavior are discussed.
\end{abstract}

\pacs{PACS numbers: 05.40.+j,64.60.Ht,05.70.Ln,68.35.Fx}

\narrowtext

\section{Introduction}

In recent years there has been much theoretical interest in the dynamics of 
growing interfaces in random media \cite{growth,surface}. This problem has 
received particular attention not only because of its technological importance
and applications\cite{surface}, but also as the simplest non--trivial example 
of dynamic scale invariance in a nonequilibrium system. Kardar, Parisi, and 
Zhang (KPZ) have proposed a simple nonlinear Langevin equation \cite{kpz86}
which has become a widely accepted and by now well--known description of the 
macroscopic aspects of a certain class of growth processes. This equation is 
closely related to a large variety of other problems ranging from 
randomly--stirred fluids \cite{fns77} (Burgers equation), dissipative transport
\cite{dds} (the driven--diffusion equation), flame--front propagation 
\cite{kus89,ks89} (Kuramoto--Sivashinski equation), polymer physics\cite{kz87},
the dynamics of a sine--Gordon chain \cite{sg89}, and the behavior of magnetic
flux lines in superconductors \cite{terry92}.

Through a simple transformation the Burgers--KPZ equation can be mapped onto 
the statistical mechanics of directed polymers in a random medium 
\cite{kz87,dp1}. This problem seems to embody many of the features of more 
complex random systems such as spin glasses \cite{dp2}. 

Due to the simplicity of its form, accompanied by a stunning complexity of its
behavior, it has gradually taken on the role of an ``Ising model of 
nonequilibrium dynamics''. Therefore, any advances in understanding the 
behavior of the Burgers--KPZ equation may have broad 
\newline\vskip 2.18truein\noindent
impacts in both the fields
of nonequilibrium dynamics as well as disordered systems. 

Assuming that a coarse--grained interface of a $d$--dimensional substrate may 
be described by a height function $h({\bf x},t)$ with ${\bf x} \in {\cal R}^d$,
KPZ proposed that the following nonlinear Langevin equation \cite{kpz86},
\begin{equation}
   {\partial h \over \partial t} = \nu {\mbox{\boldmath $\nabla$}}^2 h + 
   {\lambda \over 2} \left({\mbox{\boldmath $\nabla$}} h\right)^2 + \eta \, ,
\label{kpz}
\end{equation}
governs the macroscopic (large--distance, long--time) dynamics of a 
stochastically growing interface. The first term in Eq.~(\ref{kpz}) represents 
the surface tension which prefers a smooth surface. The second term describes 
the tendency of the surface to locally grow normal to itself, and is entirely 
nonequilibrium in origin. The last term is a Langevin noise to mimic the 
stochastic nature of any growth process. In Eq.~(\ref{kpz}), the average growth
velocity has been subtracted so that the noise has zero mean, i.e., 
$\langle{\eta({\bf x},t)}\rangle=0$. In the simplest case, the stochasticity is
then described by an uncorrelated Gaussian noise with the second moment 
\begin{equation}
   \langle{\eta({\bf x},t)\eta({\bf x}',t')}\rangle =
    2D \delta^d ({\bf x}-{\bf x}') \delta(t-t') \, ,
\label{2nd_moment_noise}
\end{equation}
where $D$ characterizes the noise amplitude. 

The shape of the steady--state surface profile is completely characterized by 
the steady--state distribution function ${\cal P}[h({\bf x},t)]$. The leading 
(second) moment of ${\cal P}$ is the truncated two--point correlation function
\begin{equation}
   C ({\bf x},t) \equiv 
    \langle { [h({\bf x}_0+{\bf x}, t_0+t)-h({\bf x}_0,t_0)]^2 } \rangle.
\label{corrdef}
\end{equation}
In Eq.~(\ref{corrdef}), the mean value of $h$ has been implicitly subtracted, 
i.e., we use a moving reference frame such that $ \langle {\partial_t h(x,t)}
\rangle = 0$. 
Because the equation of motion (\ref{kpz}) is scale invariant\cite{kpz86}, the 
correlation function assumes the scaling form 
\begin{equation}
   C (x,t) = x^{2\chi}F(t/x^z) \, ,
\label{corr}
\end{equation}
where $x \equiv |{\bf x}|$. In Eq.~(\ref{corr}), the roughness exponent $\chi$ 
describes the scaling of the width of the interface, and the dynamic exponent 
$z$ characterizes the spread in time of disturbances on the surface. In the 
limits $y = 0$ and $y \rightarrow \infty$, the scaling function $F(y)$ becomes 
$A=const.$ and $B y^{2\chi/z}$, (with $B=const.$), respectively, thus yielding 
the usual asymptotic scaling form of the correlation function,
\begin{eqnarray} 
  C (x,t=0) &= &A x^{2\chi}    \, ,  \nonumber\\
  C (x=0,t) &= &B t^{2\chi/z}  \, .  \label{corra} 
\end{eqnarray}

In cases where the nonlinear term in Eq.~(\ref{kpz}) is not relevant, one 
recovers the linear equation of Edwards and Wilkinson\cite{ew}. There the 
exponents are known exactly, $\chi_0 = (2-d)/2$, and $z_0 = 2$, as expected 
from the ensuing simple diffusion equation. 

In the more interesting case where the nonlinear term is important, one may 
derive the exponent identity 
\begin{equation}
    \chi + z = 2 \, ,
\label{exponent_identity}
\end{equation}
which follows from the invariance of (\ref{kpz}) with respect to an 
infinitesimal tilting of the surface \cite{kpz86} ($h \to h + {\bf v} \cdot 
{\bf x}$, ${\bf x} \to {\bf x} - \lambda {\bf v} t$). This invariance is known
as ``Galilean invariance'' due to the corresponding invariance of the Burgers 
equation \cite{fns77}. Therefore, there is at most one independent exponent to
be determined. However, that last link needed to determine the exponents has 
proven to be quite elusive despite considerable effort.

In the special case of $d=1$, which is realized for example in the expansion of
a domain boundary separating the two phases of a two--dimensional Ising model, 
Eq.~(\ref{kpz}) also satisfies a fluctuation--dissipation theorem (FDT). Then 
one can show \cite{growth} that the stationary distribution function is 
\begin{equation}
   P_{\rm st}[h(x)] = 
    \exp \left\{
     - {\nu \over 2D}\int dx 
    \left( 
    {\partial  h \over \partial x} 
    \right)^2 \right\} \, .
\label{stat_distribution}
\end{equation}
Hence $\chi = 1/2$, as if the nonlinearity were absent. Given the relevance of 
nonlinearity in $d=1$, one immediately obtains $z = 3/2$ through the exponent 
identity, Eq.~(\ref{exponent_identity}).

For dimensions $d > 2$ an interface described by the Burgers--KPZ equation is 
expected, on the basis of renormalization--group arguments \cite{kpz86,fns77},
to undergo a transition from an asymptotically smooth to a rough profile as the
effective strength of the nonlinear coupling $\lambda^2 D / \nu^3$ is 
increased. The exponents in the rough phase are only known from 
numerical simulations. The most recent values\cite{ft90} seem to settle between
the Wolf--Kertesz conjecture \cite{wk87} $\chi / z = 1 / (2 d + 1 )$ and the 
one by Kim and Kosterlitz \cite{kk89} $\chi/z = 1 / ( d + 2 )$. Despite those 
results from numerical simulations, there has been very little progress towards
an analytical theory which goes beyond the original paper \cite{kpz86}, at 
least on the basis of (refined) perturbational approaches. The main obstacle 
within a systematic treatment originates from a renormalization group (RG) flow
towards a strong--coupling fixed point. Therefore, dynamic renormalization 
group (DRG) methods have had very limited success. The main qualitative results
obtained \cite{kpz86,fns77} are (i) the interface is always rough (i.e., the 
nonlinearity is relevant) for $d\le 2$ \cite{fnote1}, and (ii) there is a 
dynamic phase transition between a smooth phase (nonlinearity irrelevant, 
$\chi=\chi_0, z=z_0$) and a rough phase where the nonlinearity is again 
relevant for $d > 2$. Quantitatively, DRG provides a perturbative access to the
dynamic phase transition point via a $d=2+\epsilon$ expansion. Unfortunately, 
DRG tells us virtually nothing about the most interesting aspect, namely the 
rough phase of the interface. The method breaks down because the DRG fixed 
point describing the rough phase is a {\it strong--coupling} fixed point 
generally not accessible by any perturbational means. This problem persists 
even in $d=1$ where the exponents themselves are already known exactly. 

Recently, there has been some progress in understanding the behavior at the 
roughening transition itself, which is described by an unstable fixed point in
the DRG theory. Doty and Kosterlitz \cite{dk92} have argued that the dynamic 
exponent $z_c$ at the KPZ roughening transition equals $2$ for {\it all} 
dimensions $d \geq 2$. Their argument is based on the mapping of the 
Burgers--KPZ equation to the equilibrium model of directed polymers in a random
medium, and a standard scaling argument usually applied for glassy system, also
very similar to the derivation of hyperscaling at critical points. Basically, 
they conclude that at a finite--temperature fixed point, one expects the scale
of the free energy (which corresponds to the height in the Burgers--KPZ 
equation) to be given by the temperature, which is finite. This implies for the
roughness exponent $\chi_c = 0$ and $z_c = 2$ via the exponent identity 
Eq.~(\ref{exponent_identity}). This view is indeed supported by numerical 
simulations by Tang, Nattermann, and Forrest \cite{tnf90}. Based on the 
one--loop approximation, Nattermann and Tang \cite{nat92} also discuss the 
interesting crossover scaling behavior of the Burgers--KPZ equation.

The purpose of this paper is to go beyond the one--loop perturbation theory 
which is plagued by some artefacts, and perform a two--loop calculation, in 
order to possibly clarify which aspects are generally accessible to a DRG 
treatment, and which are not, and also to establish certain trends, which might
provide clues even beyond the two--loop approximation. We shall apply the 
field--theoretical version of the DRG in the formulation of Bausch, Janssen, 
and Wagner \cite{j76,bjw76}. Using the dimensional regularization procedure 
\cite{dreg} for a massless theory such as provided by Eq.~(\ref{kpz}), requires
a clear distinction between infrared and ultraviolet singularities, only the 
latter to be included in the renormalization constants. Also, in order to be 
able to correctly describe the exactly known behavior in $d=1$ dimensions, 
performing an $\epsilon$ expansion turns out to be a very delicate procedure. 
For $d<2$ our results indicate that such an expansion becomes inconsistent, the
strong--coupling fixed point diverging for $d \rightarrow 2$, which is very 
drastically different to the usual situations, as, e.g., in the paradigmatic 
$\phi^4$ theory \cite{amit,zinn_justin}. On the other hand, for $d>2$ the now 
unstable fixed point describing the roughening transition may be treated in the
framework of a $2 + \epsilon$ expansion. The situation here can in fact be
compared to the similar treatment of the $O(n)$ nonlinear $\sigma$ model 
\cite{nsm}, where the nonlinear coupling corresponds to the transition 
temperature, and is of order $\epsilon$, too.

During the process of completing this paper, we became aware of a similar 
field--theoretic two--loop analysis of the Burgers--KPZ equation by Sun and 
Plischke \cite{sun94}. These authors, however, apply a somewhat different 
procedure, and some of their results are clearly incompatible with ours, as 
shall be discussed below.

The outline of this paper is as follows. In Section II we establish a
field--theoretic formulation for the nonlinear Langevin equation (\ref{kpz}) 
and analyze the form of the propagators and vertices. Particular attention is 
paid to the symmetries of the Burgers--KPZ equation. We discuss the 
time--reversion symmetry and the existence of fluctuation--dissipation theorems
due to detailed--balance properties of the underlying stochastic equation of 
motion. Furthermore, the Ward--Takahashi identities resulting from the Galilean
invariance of the Burgers--KPZ equation are discussed. In Section III we 
proceed with a detailed description of our renormalization procedure and 
regularization scheme, elaborating especially on how we handle the $d$ 
dependences of various origins appearing in the singular terms. We then discuss
the results of our two--loop perturbation theory, i.e., the behavior of 
Wilson's flow functions, the fixed points of the RG, and the resulting critical
exponents as functions of the dimension $d$. We shall also investigate the
scaling behavior at the dynamic roughening transition, as well as some features
of the crossover from Gaussian to critical behavior in the flat phase. 
Finally, in Section IV we discuss our results and provide an outlook on future 
developments. In the Appendices A and B we present some details of the 
two--loop calculation, listing the Feynman diagrams and analytical results for 
the two--point vertex functions, and the formulae for evaluating the integrals 
in the dimensional regularization scheme.

\section{Model equations, perturbation theory, and Ward identities}

In this section we establish a path integral formulation for the nonlinear 
Langevin equation Eq.~(\ref{kpz}), where particular attention is paid to the 
symmetries of the Burgers--KPZ equation. Readers who are not familiar with 
these techniques, may proceed to Section III.A, where a formulation of the 
dynamic renormalization group is given which does not utilize the more formal 
methods described in this section. We note, however, that the formulation of 
dynamics in terms of path integrals provides a quite powerful tool which allows
a systematic analysis of higher order terms as well as general theorems, such 
as fluctuation--disspation theorems and Ward--Takahashi identities.

\subsection{The Model and Dynamic Functional}

We begin by reformulating the stochastic dynamics in terms of path integrals.
Let us consider a coarse--grained interface of a $d$--dimensional substrate,
described by a height function $h({\bf x},t)$ with ${\bf x} \in {\cal R}^d$,
whose time evolution is governed by the Burgers--KPZ equation (the subscripts 
``0'' will henceforth denote unrenormalized quantities)
\begin{equation}
   {\partial h({\bf x},t) \over \partial t} = \nu_0 \, {\mbox{\boldmath 
    $\nabla$}}^2 h({\bf x},t)
    + {\lambda_0 \over 2} \, 
      \left({\bf {\mbox{\boldmath $\nabla$}}} h({\bf x},t)\right)^2
    + \eta({\bf x},t) \, .
\label{1}
\end{equation}
The random forces $\eta({\bf x},t)$ can be taken to be Gaussian distributed,
\begin{equation}
   W[\eta] \propto \exp 
 \left\{ - {1 \over 4 D_0} \int d^dx \int dt \; \eta({\bf x},t)^2 \right\} \, ,
\label{2}
\end{equation}
with mean zero and short--ranged spatial and temporal correlations
\begin{eqnarray}
   &&\langle \eta({\bf x},t) \rangle = 0 \, ,
 \label{3} \\
   &&\langle \eta({\bf x},t) \eta({\bf x}',t') \rangle =
      2 D_0 \, \delta^{(d)}({\bf x} - {\bf x}') \, \delta(t - t') \, .
 \label{4}
\end{eqnarray}

>From the standard dynamic field theory formulation\cite{j76,bjw76,dd75,msr73}
of Langevin dynamics, upon averaging over the noise distribution and 
introducing auxiliary response fields ${\tilde h} ({\bf x},t)$, one can turn 
the stochastic differential equation Eq.(\ref{1}) into a ``free energy'' 
functional (generating functional),
\begin{eqnarray}
   Z[j,{\tilde j}] = &&\int  {\cal D}[h] {\cal D}[i {\tilde h}] 
   \exp \Biggl\{ {\cal J}[{\tilde h}, h] \nonumber \\
   &&+         \int d^dx \int dt 
               \left[ {\tilde j} {\tilde h}+
                      j h
               \right] 
        \Biggr\}  \, ,     
\label{5}
\end{eqnarray}
with the Janssen--De Dominics functional given by
\begin{eqnarray}
{\cal J}[{\tilde h}, h] = 
   &&\int d^dx \int dt \Biggl\{ D_0 \, {\tilde h}{\tilde h}
   \nonumber \\
   &&- {\tilde h} \left[ {\partial h \over \partial t} - \nu_0 \, 
    {\mbox{\boldmath $\nabla$}}^2 h - {\lambda_0 \over 2} \, 
    ({\bf {\mbox{\boldmath $\nabla$}}} h)^2 \right] \Biggr\} \, .
\label{6}
\end{eqnarray}
Correlation and response functions can now be expressed as functional averages
with weight $\exp \left\{ {\cal J} [{\tilde h},h] \right\}$.

\subsection{Perturbation Theory}

The propagators and correlators are determined by the harmonic (``Gaussian'')
part 
\begin{eqnarray}
    &&{\cal J}_0 [{\tilde h},h] = 
    \int_{{\bf k}} \int_\omega 
    {\tilde h} ({\vec k}) \biggl\{
    D_0 {\tilde h} (-{\vec k})-
    \left[ i \omega + \nu_0 k^2 \right]
    h (-{\vec k}) \biggr\} \nonumber \\
    =&&\qquad
    -{1 \over 2} \int_{{\bf k}} \int_\omega 
    \left( {\tilde h} ({\vec k}), h({\vec k}) \right)
    {\bf A} ({\vec k})
    \pmatrix{  {\tilde h} (-{\vec k}) \cr
                       h  (-{\vec k}) \cr}
    \quad \quad \label{7}
\end{eqnarray}
of the dynamic functional, with
\begin{equation}
{\bf A} ({\vec k}) = \pmatrix{ -2 D_0 & i \omega + \nu_0 k^2 \cr
              - i \omega + \nu_0 k^2 & 0 \cr}
\label{7.a}
\end{equation}
Here we have introduced ${\vec k} = ({\bf k}, \omega)$ and defined the 
Fourier--transformed quantities by
\begin{equation}
   h({\bf x},t) = \int_{\bf q} \int_\omega h({\bf q},\omega) 
                   e^{i ({\bf q} {\bf x} - \omega t)} \quad ,
\label{8a}
\end{equation}
using the abbreviations $\int_{\bf q} ... = (2 \pi)^{-d} \int d^dq ...$ and
$\int_\omega ... = (2 \pi)^{-1} \int d\omega ...$. The interaction part 
${\cal J}_{\rm int}$ is proportional to the coupling constant $\lambda_0$ and 
given by
\begin{equation}
    {\cal J}_{\rm int} [{\tilde h},h] = 
    i^2 {\lambda_0 \over 2} 
    \int_{ \{ {\vec k}_i \} } 
    \left( {\bf k}_2 \cdot {\bf k}_3 \right)
    {\tilde h} ({\vec k}_1)
    h ({\vec k}_2)
    h ({\vec k}_3) 
   \delta (\Sigma {\vec k}_i).
\label{8b}
\end{equation}
The corresponding diagrammatic representation of the response and correlation 
propagators and the vertex is shown in Fig.~1. With the above perturbation 
theory at hand we can now calculate the cumulants $\langle  ...  \rangle _c$ of
the correlation and response functions defined by functional derivatives of 
$F [{\tilde j}, j] = \ln Z[{\tilde j},j]$ with respect to the sources 
${\tilde j}$ and $j$, respectively:
\begin{eqnarray}
  &&G_{{\tilde N},N}
  ( {\vec k}_1;...;
    {\vec k}_{\tilde N};
    {\vec k}_{{\tilde N}+1};...;
    {\vec k}_{{\tilde N}+N} )
    = \nonumber \\
    &&\qquad =
    \langle 
    {\tilde h} ({\vec k}_1) \, ... \, 
    {\tilde h} ({\vec k}_{\tilde N})
    h ({\vec k}_{{\tilde N}+1}) \, ... \,
    h ({\vec k}_{{\tilde N}+N})
    \rangle_c = \nonumber \\
    &&\qquad ={\delta^{{\tilde N}+N} \,  \ln Z [{\tilde j},{j}] \over 
              \delta {\tilde j} (-{\vec k}_1) \, ... \, 
              \delta j (-{\vec k}_{{\tilde N}+N})}
      \bigg \vert_{{\tilde j},{j} = 0} \, . 
\label{9}
\end{eqnarray}
It is convenient to consider the vertex functions $\Gamma_{{\tilde N},N}$, 
which can be obtained from the cumulants by a Legendre transformation,
\begin{equation}
  \Gamma [{\tilde h},h] = - F  [{\tilde j},j] +
  \int d^d x \int dt ( {\tilde h} {\tilde j} + h j ) \, ,
\label{10}
\end{equation}
where
\begin{equation}
  h = { \delta F \over \delta j}, \quad \quad 
  {\tilde h} = {\delta F \over \delta {\tilde j}} \, .
\label{11}
\end{equation}
In equilibrium dynamics the linear response is defined by adding to the 
Hamiltonian an external field  which couples linearly to the field $h$. For 
nonequilibrium random processes, however, the stationary probability 
distribution function (which is the analogon of the Gibbsian measure) is not 
known a priori, except for certain cases where the potential conditions (see 
below) are fulfilled and the random process obeys a detailed--balance 
condition \cite{dh75,g73}. Therefore, we define the linear response in terms of
(deterministic) ``external'' fields $b$ added linearly to the drift term in the
equation of motion
\begin{equation}
   V(h) \rightarrow V(h) + b.
\label{12}
\end{equation}
This ``external'' field $b$ corresponds to the source ${\tilde j}$ in the 
generating functional. This remark should also clarify why ${\tilde h}$ is 
called a ``response field'' (see Ref.~\cite{bjw76}).

We proceed with a simple dimensional analysis to determine the engineering 
dimensions of the parameters and fields. Since the dynamic functional has to be
dimensionless $[{\cal J}] = 1$, upon defining inverse length and time scales 
according to $[q] = \Lambda$ and $[\omega] = \nu_0 \, \Lambda^2$, respectively,
we arrive at the following ``naive'' dimensions
\begin{eqnarray}
   &&[h] = \Lambda^{-d/2-3} \, \nu_0^{-3/2} \, D_0^{1/2} \, ,
 \label{13} \\
   &&[{\tilde h}] = \Lambda^{-d/2-1} \, \nu_0^{-1/2} \, D_0^{-1/2} \, ,
 \label{14} \\
   &&[\lambda_0] = \Lambda^{-d/2+1} \, \nu_0^{-3/2} \, D_0^{-1/2} \, .
 \label{15}
\end{eqnarray}
Hence the engineering dimension of the effective coupling constant is 
$[\lambda_0^2 D_0 / \nu_0^3] = \Lambda^{2-d}$. The space dimension at which the
effective coupling constant becomes dimensionless is $d_c = 2$. In simple cases
such as a $\phi^4$ theory this is the dimension which separates Gaussian from 
critical behavior. However, this simple scenario is not at all what one finds 
for the Burgers--KPZ equation. As we will see in the following, $d_c=2$ is the 
dimension below which the behavior of the interface fluctuations are described
by a finite ``strong coupling'' fixed point, which is {\it not} of order 
$\epsilon = d-2$, but rather of order ${\cal O}(1)$. Above $d_c=2$, on the 
other hand, the two--loop calculation yields an infrared unstable fixed point 
which {\it is} of order ${\cal O}(\epsilon)$. For effective coupling constants 
less than this unstable fixed point, the RG flow tends to small coupling, i.e.,
the interface fluctuations are described by a capillary wave Hamiltonian 
${\cal H} \propto \int d^dx ({\mbox{\boldmath $\nabla$}} h)^2$ (corresponding 
to ``Gaussian'' behavior). When the coupling constant becomes larger than the 
value of the unstable fixed point, the RG flow tends to infinity. The 
corresponding {\it strong--coupling} behavior is hence described by a 
``strong--coupling'' fixed point which is obviously not accessible by 
perturbative methods (of any finite order in perturbation theory). Resummation
techniques such as the quite successful mode--coupling approach 
\cite{mode_coupling} are needed. Those methods have been applied to the 
Burgers--KPZ problem in Refs.~\cite{dds,mode_kpz,erwin,bc93,dmkb94,helge,tu94}.
They allow a determination of the critical exponents and the scaling functions.
However, little is known to date, whether or in what sense those 
self--consistency methods constitute a controlled expansion with some possibly
small parameter. The two--loop calculation in this paper provides some hints on
the validity of the mode coupling theory in the ($1+1$)--dimensional case (see 
Section III).

\subsection{Ward--Takahashi Identities}

As usual, internal symmetries of the effective Lagrangian, i.e. here: the 
Langevin equation, are important in determining the number of {\it independent}
renormalization factors of the theory. The most important symmetry of the 
Burgers--KPZ equation is its behavior with respect to tilting of interface by 
an (infinitesimal) angle ${\bf v}$ (this corresponds to the Galilean invariance
of Burgers' equation in hydrodynamics)
\begin{eqnarray}
   h'({\bf x},t) &&= h({\bf x} + \lambda_0 {\bf v} t, t) 
                     + {\bf v}  \cdot {\bf x} \, ,
 \label{16} \\
   {\tilde h}'({\bf x},t) &&= 
             {\tilde h}({\bf x} + \lambda_0 {\bf v} t, t) \, .
 \label{17}
\end{eqnarray}
In Fourier space the latter equations read
\begin{eqnarray}
   h'({\bf q},t) &&= e^{i \lambda_0 {\bf v} \cdot {\bf q} t} 
                     h({\bf q},t)
                     - i {\bf v} \cdot
                         {\partial \over \partial {\bf q}}
                       \delta({\bf q}) \, ,
 \label{18} \\
   {\tilde h}'({\bf q},t) &&= 
         e^{i \lambda_0 {\bf v} \cdot {\bf q} t} 
         {\tilde h}({\bf q},t) \, .
 \label{19}
\end{eqnarray}
The invariance of the generating functional of the vertex functions under the 
above transformations implies the following Ward--Takahashi identity
\begin{eqnarray}
   \delta \Gamma  = {\bf v} \cdot \int_{\bf q} \! \int \! dt 
    \Biggl[ &&i \lambda_0 {\bf q} t \left( 
     {\delta \Gamma \over \delta h({\bf q},t)} h({\bf q},t) + 
     {\delta \Gamma \over \delta {\tilde h}({\bf q},t)} {\tilde h}({\bf q},t)
    \right) \nonumber \\
    &&+ i \delta({\bf q}) {\partial \over \partial {\bf q}} 
     {\delta \Gamma \over \delta h({\bf q},t)} \Biggr] = 0 \, .
 \label{20}
\end{eqnarray}
Functional derivatives with respect to ${\tilde h}({\bf q}',t')$ and 
$h({\bf q}'',t'')$, ${\tilde h}({\bf q}'',t'')$ yield relations between 
three--point and two--point vertex functions
\begin{eqnarray}
    \lambda_0 \, ({\bf q}' t' &&+ {\bf q}'' t'') 
    \Gamma_{{\tilde h} h}({\bf q}',t';{\bf q}'',t'') = \nonumber \\ =&&
     - {\partial \over \partial {\bf q}} \int dt 
      \Gamma_{{\tilde h} h h}({\bf q}',t';{\bf q}'',t'';{\bf q},t) 
       \bigg \vert_{{\bf q} = 0} \, ,
 \label{21} \\
    \lambda_0 \, ({\bf q}' t' &&+ {\bf q}'' t'') 
    \Gamma_{{\tilde h} {\tilde h}}({\bf q}',t';{\bf q}'',t'')
    = \nonumber \\ =&&
     - {\partial \over \partial {\bf q}} \int dt 
      \Gamma_{{\tilde h} {\tilde h} h}({\bf q}',t';{\bf q}'',t'';{\bf q},t) 
       \bigg \vert_{{\bf q} = 0} \, ,
 \label{22}
\end{eqnarray}
or, in frequency representation
\begin{eqnarray}
    i \lambda_0 \, ({\bf q} &&\cdot  {\bf p}) {\partial \over \partial \omega}
    \Gamma_{{\tilde h} h}({\bf q},\omega;-{\bf q},-\omega) = \nonumber \\ =&&
     \Gamma_{{\tilde h} h h}({\bf q}-{\bf p},\omega;
                             -{\bf q},-\omega;{\bf p},0) \, ,
 \label{23} \\
   i \lambda_0 \, ({\bf q} &&\cdot {\bf p}) {\partial \over \partial \omega}
    \Gamma_{{\tilde h} {\tilde h}}({\bf q},\omega;-{\bf q},-\omega)
    = \nonumber \\ =&&
     \Gamma_{{\tilde h} {\tilde h} h}({\bf q}-{\bf p},\omega;
                                      -{\bf q},-\omega;{\bf p},0) \, .
 \label{24}
\end{eqnarray}

>From the diffusive dynamics and the corresponding ${\bf q}$ dependence of the 
vertices follows the {\it exact} result (valid to any order in perturbation 
theory)
\begin{equation}
   \Gamma_{{\tilde h} h}({\bf 0},\omega) = i \omega \, .
 \label{25}
\end{equation}
Hence the fields ${\tilde h}$ and $h$ do not renormalize\cite{fnote2}. The Ward
identies, Eqs.~(\ref{23}) and (\ref{24}), then yield immediately that there is
also no renormalization of the nonlinearity $\lambda$. Therefore we shall 
henceforth drop the index ``0'' for $\lambda$, previously denoting the 
unrenormalized quantity, as we have already anticipatingly done for the fields.

\section{Renormalization Group Theory}

In this section we present our results from the renormalization group analysis
of the Burgers--KPZ equation to two--loop order. We focus on the main concepts 
and results. For more details we refer the reader to the appendices, where some
intermediate results for the calculations of the vertex functions and the 
renormalization constants are presented.

\subsection{General Considerations}

By now several powerful methods have evolved, and are frequently used, by which
dynamical critical phenomena are analyzed. The method by which the 
Burgers--KPZ equation has been investigated first \cite{fns77}, is based on the
dynamic renormalization group as formulated by Ma and Mazenko \cite{mama75}. 
This scheme is in close analogy to Wilson's momentum--shell procedure 
\cite{wiko74}, and has the appealing feature of being conceptually transparent 
and more ``physical'' than field--theoretic methods, which utilize a mapping of
the stochastic equations of motion onto a dynamical functional 
\cite{msr73,dd75,j76,bjw76}, and concepts originally developed for 
understanding quantum field theories \cite{amit,zinn_justin}. The latter 
schemes have the advantage of providing a powerful tool for calculating 
correlation functions and interconnections between them in a systematic way, 
which can be crucial if one intends to go beyond one--loop order.

Since the two--loop calculations will become quite complicated, and there is 
certainly the danger that the important physical and conceptual points may be
obscured by the tedious calculations, we regard it as useful at this point to 
review the dynamic renormalization group procedure applying both Wilson's 
scheme and field--theoretic methods, to one--loop approximation. In order to 
keep our arguments as simple as possible, we also refrain from using the 
mapping onto a dynamic functional in this chapter, and formulate the 
theoretical concepts by using just the equation of motion.

Eq.~(\ref{kpz}) reads in Fourier space
\begin{eqnarray}
  h({\vec q}) = &&G_0({\vec q}) \eta({\vec q}) \nonumber \\
                     + &&{1 \over 2} G_0({\vec q})
                          \int_{{\vec k}} V_0({\bf q}_+;{\bf q}_-)
                          h({\vec q}_+) h({\vec q}_-)  \, . 
\label{3.1}
\end{eqnarray}
In Eq.~(\ref{3.1}) $G_0({\bf q},\omega) = 1/(\nu_0{\bf q}^2 -i\omega)$ denotes 
the ``bare propagator'', $V_0({\bf q}_1;{\bf q}_2) = - \lambda_0 \, {\bf q}_1 
\cdot{\bf q}_2$ is the ``bare vertex'', and ${\bf q}_\pm \equiv {\bf q}/2 \pm 
{\bf k}$, $\omega_\pm \equiv \omega/2 \pm \nu$. For $V_0=0$, Eq.~(\ref{3.1}) is
just the linear diffusion equation. For $V_0 \neq 0$, the solution of 
Eq.~(\ref{3.1}) may be obtained iteratively by a perturbation expansion in 
powers of $V_0$. For example, the lowest--order correction to the response 
function $G = \delta \langle h \rangle / \delta \eta$ is
\begin{equation}
   G_1({\vec q}) 
   = G_0({\vec q}) + 
     G_0({\vec q}) \Sigma_1({\vec q}) G_0({\vec q}) \, ,
\label{3.2}
\end{equation}
where $C_0({\vec q}) = 2 D_0 |G_0({\vec q})|^2$ is the ``bare 
correlator'', and
\begin{equation}
   \Sigma_1({\vec q}) = - \int_{\vec k} 
      V_0({\bf q}_+;{\bf q}_-) 
      G_0({\vec q}_-) C_0({\vec q}_+) 
      V_0({\bf q}_+;{\bf q})
\label{3.3}
\end{equation}
represents the one--loop renormalization of the ``self--energy''. Similarly, 
the lowest--order correction to the correlation function is
\begin{equation}
  C_1({\vec q}) = 2 D_1({\vec q}) \mid G_0({\vec q}) \mid^2, 
\label{3.4}
\end{equation}
where
\begin{equation}
    D_1({\vec q}) = D_0 + {1 \over 4} \int_{{\vec k}}  
       V_0({\bf q}_+;{\bf q}_-) C_0({\vec q}_-) C_0({\vec q}_+)
       V_0({\bf q}_+;{\bf q}_-)
\label{3.5}
\end{equation}
is the one--loop renormalization of the ``noise spectrum''. Higher loop orders 
could in principle be obtained from iteration of the equation of motion, 
Eq.~(\ref{3.1}). But, since we are mainly interested in the conceptual problems
now, we restrict ourselves for the sake of simplicity to the one--loop 
approximation. (The two--loop contributions and their corresponding 
diagrammatic representations are presented in Appendix A and will be discussed
in Subsection B.) After performing the internal frequency integrals one obtains
for the response and correlation functions at zero frequency $\omega = 0$ and 
in the limit ${\bf q} \rightarrow {\bf 0}$, respectively, the following results
\begin{eqnarray}
 C_1(&&{\bf 0},0)
  = C_0({\bf 0},0) \nonumber \\ 
  + &&2 {\lambda^2 D_0 \over 4 \nu_0^3} G_0({\bf 0},0)^2 
     \int {d^d k \over (2 \pi)^d} {1 \over k^2} + {\cal O}(q^2) \, ,
\label{3.6} \\
  G_1(&&{\bf q},0)
  = G_0({\bf q},0) \nonumber \\
  - &&{\lambda^2 D_0 \over 4 \nu_0^3} G_0({\bf q},0)^2 
   {2-d \over d} q^2 \int {d^d k \over (2 \pi)^d} {1 \over k^2}
  + {\cal O}(q^4) \, ,
\label{3.7}
\end{eqnarray}
where only the terms of lowest order in $q^2$ have been retained. It is 
essential to note that the prefactor $2-d$ in Eq.~(\ref{3.7}) is a {\it 
geometry} factor, stemming from the scalar products in the integrand, i.e., we 
have made use of the identity $\int_{\bf k} ({\bf q} \cdot {\bf k}) f(k) = 
(q^2/d) \int_{\bf k} k^2 f(k)$ with $k = \mid {\bf k} \mid$. This 
$d$--dependent prefactor has nothing at all to do with the problems in 
perturbation theory, resulting from infrared divergences, to be discussed next.
Namely, the ``naive'' perturbation theory breaks down due to the fact that the 
integral $\int_{\bf k} k^{-2}$ in Eqs.~(\ref{3.6}) and (\ref{3.7}) is 
manifestly {\it infrared (IR)--divergent} at the lower cutoff. These 
divergences are of real physical origin, and it is exactly those, by which one 
is lead to the introduction of the renormalization--group concept. 

In {\it Wilson's momentum space procedure} \cite{wiko74} the RG transformation 
is defined as follows:
\par
\noindent (1) {\it Elimination}: First, one eliminates those modes within a 
small momentum shell $k \in [\Lambda/b,\Lambda]$, where $b = e^l > 1$. This 
results in an effective (renormalized) propagator for the remaining modes and 
allows the identification of an effective surface tension (diffusion constant) 
to one--loop order \cite{fns77,kpz86}
\begin{equation}
 \nu_{\rm w} = \nu_0 \left[1 + {\lambda^2 D_0 \over 4 \nu_0^3} 
         \left( {2-d \over d} \right)  K_d \Lambda^{d-2} l \right] \, ,
\label{3.8}
\end{equation}
where $K_d = S_d / (2 \pi)^d$ and $S_d$ is the surface area of the unit sphere 
in $d$ dimensions. In an analogous way one finds \cite{fns77,kpz86}
\begin{equation}
 D_{\rm w} = D_0 \left[1 + {\lambda^2 D_0 \over 4 \nu_0^3} 
         \left( {1 \over d} \right)  K_d  \Lambda^{d-2} l  \right] \, .
\label{3.9}
\end{equation}
Note that the expressions (\ref{3.8}) and (\ref{3.9}) coincide in $d=1$, as is
required by the additional fluctuation--dissipation theorem (FDT) valid only 
at this specific dimension. Without loss of generality we may set the cutoff 
$\Lambda$ to unity.

\par
\noindent (2) {\it Rescaling:} After this elimination--step has been performed,
the resulting model has a cutoff which is reduced by a factor $b=e^{l}>1$. In 
order to remove this difference and arrive at a form more closely resembling 
the original system, the momenta, times, and height variable are rescaled 
according to $k \rightarrow k e^{-l}$, $t \rightarrow e^{zl} t$, and $h 
\rightarrow e^{\chi l} h$, respectively. Upon requiring that the equation of 
motion stays invariant under this scale transformation, the parameters and 
coupling constants transform as follows
\begin{eqnarray}
 \nu_0 \rightarrow && b^{z-2} \nu_0 \, , \label{3.10} \\
 D_0 \rightarrow &&b^{-d - 2 \chi + z} D_0 \, , \label{3.11} \\
 \lambda_0 \rightarrow &&b^{\chi + z - 2} \lambda_0 \, .
\label{3.12}
\end{eqnarray}

Combining these contributions results in differential recursion relations (for 
infinitesimal RG transformations where $l \ll 1$) \cite{fns77,kpz86,nat92}:
\begin{eqnarray}
 {d \nu_{\rm w} \over dl} = &&\nu_{\rm w} \left[ z - 2 + K_d {\lambda_{\rm w}^2
 D_{\rm w} \over 4 \nu_{\rm w}^3} \left( {2-d \over d} \right) \right] \, , 
\label{3.13} \\
 {d D_{\rm w} \over dl} = &&D_{\rm w} \left[ z - 2 \chi - d + K_d 
 {\lambda_{\rm w}^2 D_{\rm w} \over 4 \nu_{\rm w}^3} \right] \, ,
\label{3.14} \\
 {d \lambda_{\rm w} \over dl} = &&\lambda_{\rm w} \left[ \chi + z - 2 \right] 
 \, ,  \label{3.15}
\end{eqnarray}
where the suffix ``w'' indicates that the running coupling constants and 
parameters are obtained within the framework of Wilson's momentum shell method.
(We shall later introduce another set of flow functions in the context of
renormalized field theory.) Note that there is no perturbative correction to 
the flow of $\lambda_{\rm w}$, which can be understood as a direct consequence
of the Ward--Takahashi identities (\ref{21})--(\ref{24}).

\par
\noindent (3) {\it Fixed point and critical exponents:} 
The last step is to determine the exponents $z$ and $\chi$ such that the 
parameters $\nu_{\rm w}$ and $D_{\rm w}$ are not changed, for we have then
obviously arrived at a scale--invariant situation. We define an effective
coupling constant by 
\begin{equation}
   g_{\rm w} = K_d {\lambda_{\rm w}^2 D_{\rm w} \over 4 \nu_{\rm w}^3} \, ,
\label{3.15a}
\end{equation}
and obtain for the corresponding flow:
\begin{equation}
  {d g_{\rm w} (l) \over dl} =   {(2 - d)} g_{\rm w} (l) + 
  {2(2 d - 3) \over d} g_{\rm w}^2 (l) + {\cal O}(g_{\rm w}^3) \, .
\label{3.16}
\end{equation}
>From these recursion relations one can solve for the RG fixed points, i.e., 
points in parameter space which are invariant under scale transformations. Thus
setting $d g_{\rm w}(l)/dl = 0$ in Eq.~(\ref{3.16}) yields the ``Gaussian'' 
fixed point $g_0^* = 0$, and also one non--trivial fixed point
\begin{equation}
  g_1^* = {d (d-2) \over 2 (2d-3)} \, ,
\label{3.16a}
\end{equation}
which is positive for $0 \leq d < 3/2$, diverges at $d = 3/2$, then rises again
from minus infinity, crossing the zero axis at $d=2$, and becoming positive
for $d > 2$ (Fig.~2). In the physical region, $g > 0$, $g_1^*$ turns out to be
stable (``attractive'') for $d < 3/2$, thus providing non--trivial 
(strong--coupling) scaling behavior, while for $d > 2$ the Gaussian fixed point
is stable. Hence, depending on its initial value, $g_{\rm w}(l)$ will either 
flow to $g_0^*$, or to the strong--coupling ``fixed point'' $g_{\rm w} 
\rightarrow \infty$. Thus for $d > 2$  a dynamic phase transition (``roughening
transition'') is found, governed by the unstable fixed point $g_1^*$. The fact
that there is no finite positive fixed point in the range $3/2 < d < 2$ will 
turn out to be an artefact of the one--loop approximation (see Subsection B).

Upon setting $d \nu_{\rm w} / dl = 0$ and $d D_{\rm w} / dl = 0$, and 
$g_{\rm w} = g^*$, one finds $z_0 = 2$ and $\chi_0 = (2-d)/2$ at the Gaussian 
fixed point, while at any non--trivial fixed point $g^* \not= 0$ the critical 
exponents become
\begin{eqnarray}
  z = &&2 +  \left( {d-2 \over d} \right) g^* \, , 
\label{3.17}\\
  \chi = && -  \left( {d-2 \over d} \right) g^* \, . 
\label{3.18}
\end{eqnarray}
Note that Eq.~(\ref{3.15}) forces $z + \chi = 2$, if $\lambda \not= 0$. Thus
the exponent identity (\ref{exponent_identity}) holds for {\it any finite}
fixed point, which implies that knowledge of either the flow of $\nu_{\rm w}$
or $D_{\rm w}$ suffices in order to determine the critical exponents in its
vicinity.

(As already mentioned, the flow obtained from these one--loop equations still 
contains some artefacts, which are due to the low order of perturbation theory
considered. Yet, our concern in this chapter is rather a comparison of the 
various methods available to determine the RG flow of the parameters and 
couplings than a detailed description of the physical behavior.)

Let us now compare these results with the various field theoretic techniques 
using different regularization procedures. In field theory, the cut--off 
$\Lambda$ is set to infinity, thus leading to ultraviolet (UV) divergences 
above a certain critical dimension $d_c$ (in our case, $d_c = 2$). There are 
basically two regularization methods, by which these integrals are assigned 
meaningful values, that are frequently used in field theories: (i) 
reintroducing the cut--off, and (ii) dimensional regularization. In the former
case the original UV singularities appear as logarithms or powers of $\Lambda$,
in the latter as $1/(d_c-d)$ poles. The criteria of choosing either of them are
mostly guided by convenience, i.e., the amount of effort by which the 
corresponding integrals may be evaluated. Using those more formal 
field--theoretic methods one has to be careful, however, especially if one 
deals with massless theories, as is the case here! Let us explain why some 
specific attention is required (see also Ref. \cite{zinn_justin}, p.161). In 
dimensional regularization \cite{dreg,amit} one has
\begin{equation}
  \int {d^d p \over p^m} = 0 \, ,
\label{3.19}
\end{equation}
which is a (non--trivial for even $m$) consequence of the dilatation property 
of the integrals defined here by dimensional continuation. This result can be 
regarded as a cancellation between infrared (IR) and ultraviolet (UV) 
divergences. This often convenient property of integrals within the dimensional
regularization scheme may, however, have quite dangerous consequences: In a 
field theory involving {\it massless} fields (for which the Burgers--KPZ 
equation is an important example), the latter generate IR singularities, which,
again, have the signature of $1/(d_c-d)$ poles! Thus, contrary to calculations
using a finite cut--off, the dimensional regularization scheme is in danger of
mixing UV and IR poles, as is obvious from Eq.~(\ref{3.19}). {\it It is,
however, crucial to clearly separate the UV divergences, and avoid any mixing
with the IR singularities,} as will be explained below. This really may be the
source of many mistakes, if this method is applied without sufficient 
precaution. Thus it is {\it essential} to employ an IR cut--off in the 
integrals. In the case of the Burgers--KPZ equation, we cannot simply introduce
a mass term, however, because this would immediately violate the Galilean
invariance discussed in Section II.C. Therefore we have to evaluate all
quantities explicitly at {\it finite} external wave vectors or frequencies. For
the calculations in Appendix B, we have chosen the convenient normalization 
point NP: ${\bf q} = 0$, $i \omega / 2 \nu = \mu^2$.

In order to make our arguments clear we want to very briefly review some of the
fundamental ideas behind renormalized field theory and the connection to 
critical phenomena \cite{amit,zinn_justin,itz_drouffe}. From the dimensional 
engineering in Section II and simple power counting arguments, it follows that 
perturbation theory fails to describe the critical theory below the critical 
dimension $d_c$ because of IR divergences (see above). Above $d_c$, on the 
other hand, the integrals become UV divergent. These UV singularities may be 
handled very effectively by using the well--established technology of quantum 
field theory, namely by introducing renormalization constants into which the UV
poles are absorbed. From these the Callan--Symanzik equations can be inferred, 
which describe the dependence of the renormalized quantities on the 
renormalization scale. Already by a heuristic argument (see Ref. 
\cite{itz_drouffe}, p. 271), the intimate relation of the IR and UV 
singularities may be explained as follows. As mentioned above, we shall have to
carefully employ a finite IR cut--off, say a mass parameter $\mu$ (which in 
our case is rather the external frequency). The physical IR divergences in 
which we are primarily interested, manifest themselves in the domain $q / \mu 
\gg 1$, and $q / \Lambda \ll 1$, where ${\bf q}$ are the external momenta. By 
construction the latter condition is automatically satisfied in renormalized 
field theory. The former can also be read as $q \rightarrow \infty$ for fixed 
$\mu$. Note that such a scaling argument is possible, only when the 
loop--integrals are arranged in such a way that they are insensitive to their 
upper limits (this is what one does in renormalizing the theory). More 
specifically (and technically), a change in the normalization point $\mu 
\rightarrow b \mu$ of the theory may be interpreted as a scale transformation 
$q \rightarrow q/b$, and solving the Callan--Symanzik equation provides the 
corresponding RG flow equations. Again the fixed points with respect to these 
(infinitesimal) scale transformations are investigated, and the fixed point 
values of Wilson's functions, the so--called ``anomalous dimensions'', yield 
the critical exponents.

Let us explain this for the case of the Burgers--KPZ equation. In order to 
renormalize the theory (i.e., remove the UV divergences) one introduces 
renormalized parameters, which are related to the ``bare'' parameters through
renormalization factors containing all the UV poles. Before turning to the
one--loop results, some general features of the renormalization, which are 
valid to every order in perturbation theory, should be noted. As discussed in 
the previous section, the Ward--Takahashi identities 
Eqs.~(\ref{21})--(\ref{24}) together with Eq.~(\ref{25}) ensure that the 
renormalization involves only two independent renormalization factors, namely,
the renormalization of the noise amplitude $D_0$ and the surface tension 
$\nu_0$. Hence we define renormalized parameters by
\begin{equation}
   D = Z_D \, D_0 \quad , \qquad \nu = Z_\nu \, \nu_0 \, ,
 \label{3.20}
\end{equation}
and determine these by the following normalization conditions for the singular 
parts of the two--point vertex functions
\begin{eqnarray}
   \Gamma_{{\tilde h} {\tilde h}}({\bf q},\omega) 
     \vert_{\rm NP}^{\rm sing} &&= - 2 D \, ,
 \label{3.21} \\
   {\partial \over \partial q^2} \Gamma_{{\tilde h} h}({\bf q},\omega) 
     \vert_{\rm NP}^{\rm sing} &&= \nu \, .
 \label{3.22}
\end{eqnarray}
The normalization point (NP) is conveniently chosen at ${\bf q} = {\bf 0}$ and
$i \omega / 2 \nu = \mu^2$, and $1/\mu$ is an arbitrary length scale. We remark
that this is {\it not} a minimal--subtraction prescription, as was applied by 
Sun and Plischke \cite{sun94}, where just the residues of the $1/(d_c-d)$ poles
(in the dimensional regularization scheme) would be included in the Z factors.
Rather we retain the complete dependence on the dimension $d$ in the geometric
prefactors originating in the angular integrations [see the discussion 
following Eqs.~(\ref{3.6}) and (\ref{3.7})].

The corresponding Wilson functions $\zeta_\nu$, $\zeta_D$ and the $\beta$ 
function $\beta_g$ (see below) for the effective coupling constant
\begin{equation}
   g_0 = {\lambda_0^2 D_0 \over 4 \nu_0^3}
 \label{3.23}
\end{equation}
permit us to study the renormalization--group flows of the renormalized 
parameters and coupling constant $D$, $\nu$ and $g$. Note again that due to the
Ward--Takahashi identities there is no renormalization for the amplitude 
$\lambda$ of the three--point vertex to all orders of $g_0$, i.e., $Z_\lambda =
1$. Hence one gets for the renormalized effective coupling constant $g$
\begin{equation}
   g = Z_g \, g_0 \, C_d \, \mu^{d-2} \quad {\rm with} \quad 
   Z_g = Z_D Z_\nu^{-3} \, ,
 \label{3.24}
\end{equation}
where we have absorbed the geometry factor $C_d = K_d \Gamma(d/2) \Gamma(2-d/2)
= \Gamma(2-d/2) / 2^{d-1} \pi^{d/2}$ in the definition of the renormalized 
coupling.

To one--loop order the bare vertex functions read 
\begin{eqnarray}
  \Gamma_{{\tilde h}{\tilde h}}&& ({\bf 0}, \omega)^{sing} = \nonumber \\ 
  = &&-2 D_0 \Biggl[ 1 + {\lambda_0^2 D_0 \over 4 \nu_0^3} 
    {\rm Re} \left( \int_{\bf k} {1 \over i \omega / 2 \nu_0 + k^2} 
                        \right) \Biggr] \, , 
\label{3.25} \\
  {\partial \over \partial q^2}
   &&\Gamma_{{\tilde h}{\tilde h}} ({\bf q}, \omega)^{sing} 
   \mid_{{\bf q} = {\bf 0}}  = \nonumber \\ =&&\nu_0 \Biggl[ 
   1 - {\lambda_0^2 D_0 \over 4 \nu_0^3} {d-2 \over d} 
   {\rm Re} \left( \int_{\bf k} {1 \over i \omega / 2 \nu_0 + k^2} \right)
                         \Biggr] \, ,
\label{3.26}
\end{eqnarray}
where the $d$--dependent prefactor in the latter equation is of exactly the 
same geometric origin as in Eq.~(\ref{3.7}) above. In $d=1$ the results for the
amplitude of the noise spectrum and the surface tension again become identical
as is required by the fluctuation--dissipation theorem valid in $d=1$. This
important feature would have been lost in a minimal subtraction prescription! 
~\cite{fnote3}. Applying dimensional regularization, the above integrals yield
at the normalization point $i \omega / 2 \nu = \mu^2$:
\begin{equation}
 \int_{\bf k} {1 \over \mu^2 + k^2} = - {C_d \mu^{d-2} \over d-2} \, .
\label{3.27}
\end{equation}
The $1/(d-2) = 1/\epsilon$ pole corresponds to a $\ln \Lambda$ in the 
cutoff--regularization scheme, and to the momentum shell integral $\propto l$ 
in Wilson's scheme. At this point we emphasize that this $1/\epsilon$ pole has 
to be subtracted in order to renormalize the theory (i.e., remove the UV 
divergences corresponding to $\ln \Lambda$ at the critical dimension $d_c=2$). 
Let us now demonstrate once more why the dimensional regularization method 
(with {\it minimal} subtraction) is a dangerous procedure unless carried out
with considerable precaution. Namely, one might argue that the $1/\epsilon$
pole for the noise amplitude is cancelled by the prefactor $d-2$ and hence does
not have to be incorporated in the renormalization factors \cite{sun94}. In our
opinion, that would be incorrect! This procedure would definitely leave UV 
divergences in the theory, as becomes obvious in the cut--off regularization 
scheme, where $\ln \Lambda$ terms would survive. In summary, one should keep in
mind that the $1/\epsilon$ poles in the dimensional regularization scheme are 
essentially nothing else but a quite convenient way of keeping track of the UV
($\ln \Lambda$) divergences at the critical dimension $d_c=2$ (the 
simplification is that the resulting integrals are much easier to carry out in 
the dimensional regularization method); {\it one should thus strictly avoid 
mixing these ``artificial'' dimensional factors with others originating from 
purely geometrical properties of the integrals}. Again, this might constitute a
troublesome trap unless sufficient attention is paid to this issue. Henceforth,
and in the Appendix B, we try to emphasize this by clearly distinguishing
between ``$d$'' and ``$\epsilon$''. 

Using the definitions (\ref{3.20}), the normalization conditions (\ref{3.21}),
(\ref{3.22}), and the one--loop results (\ref{3.25}) and (\ref{3.26}) for the 
singular parts of the two--point vertex functions, one arrives at
\begin{eqnarray}
  Z_D   = &&1 - {g_0 C_d \mu^\epsilon \over \epsilon} \, , 
\label{3.28} \\
  Z_\nu = &&1 + \left( {d-2 \over d} \right) 
                {g_0 C_d \mu^\epsilon \over \epsilon} \, . 
\label{3.29}
\end{eqnarray}
Upon defining Wilson's flow functions by
\begin{equation}
   \zeta_D = \mu {\partial \over \partial \mu} \Big \vert_0 \ln Z_D \, , \qquad
   \zeta_\nu = \mu {\partial \over \partial \mu} \Big \vert_0 \ln Z_\nu \, ,
\label{3.30}
\end{equation}
and the $\beta$ function
\begin{equation}
   \beta_g = \mu {\partial \over \partial \mu} \Big \vert_0 g
           = g (d-2 + \zeta_D - 3 \zeta_\nu) \, ,
\label{3.31}
\end{equation}
we find the following the one--loop results
\begin{eqnarray}
   \zeta_D = &&- g + {\cal O}(g^2) \, , \label{3.32} \\
   \zeta_\nu = && {d-2 \over d} g + {\cal O}(g^2) \, , \label{3.33} \\
   \beta_g = && g \left( d-2 - {2(2d - 3) \over d} g \right) \, . \label{3.33a}
\end{eqnarray}
Note that these ``physical'' quantities do not contain the artificial 
$\epsilon$ factors any more!

Running parameters and coupling ``constants'' are now defined by (see 
Subsection B)
\begin{eqnarray}
  l {d D \over dl} = &&\zeta_D (l) D(l) \, , \label{3.34}  \\
  l {d \nu \over dl} = &&\zeta_\nu (l) \nu(l)  \, , \label{3.35} \\
  l {d g \over dl} = &&\beta_g (l)  \, . \label{3.36} 
\end{eqnarray}
We remark that these flow equations are related to the flow equations in 
Wilson's scheme by the following replacements [see Eqs.~(\ref{3.13}), 
(\ref{3.14}), and (\ref{3.17})]:
\begin{eqnarray}
  \ln (1/l) &&\rightarrow l \, , \label{3.37} \\
  D         &&\rightarrow D_{\rm w} = D e^{(z-2\chi-d)l} \, , \label{3.38} \\
  \nu       &&\rightarrow \nu_{\rm w} = \nu e^{(z-2)l} \, . \label{3.39} 
\end{eqnarray}
By determining the zeros of the $\beta$ function (\ref{3.31}) we find the same
fixed points $g_0^*$ and $g_1^*$ as in Wilson's scheme [Eq.~(3.18), Fig.~2]; 
note, however, that we have incorporated somewhat different geometry factors in
the definitions of the renormalized coupling $g$ ({\ref{3.24}) and $g_{\rm w}$ 
({\ref{3.16}). As will be demonstrated in the next subsection, the critical 
exponents are given by $z = 2 + \zeta_\nu^*$ and $\chi = 1 - (d + \zeta_D^* - 
\zeta_\nu^*)/2$, respectively. This yields the identical results as the 
momentum shell procedure, demonstrating (to this order) that both schemes are 
equivalent methods in order to find the universal critical behavior.

A very important final remark is in place here. In either scheme, the above
one--loop results were obtained by evaluating the integrals at {\it fixed}
dimension. If an expansion near $d=2$ were applied, hardly any answer would
have been found for $d < 2$, due to the divergence of the fixed point $g_1^*$ 
at $d = 3/2$. This is precisely the regime which is addressed by Sun and 
Plischke \cite{sun94} in their recent two--loop calculation. They do indeed 
find another {\it finite} fixed point at $d=2$ with their approach, but it is
difficult to see how this result may be consistent with the implicit assumption
of the $\epsilon$ expansion that any non--trivial fixed point be of order
$\epsilon$. Our own findings within the two--loop approximation, as explained 
below, rather seem to indicate that a ``full'' $2 - \epsilon$ expansion cannot
be performed consistently, in accord with the trend already seen on the 
one--loop level. The situation is entirely different for $d > 2$. Here one may 
follow the ideas exploited in the study of the $O(n)$ nonlinear $\sigma$ model 
\cite{nsm}, and perform an $\epsilon$ expansion {\it above} the critical 
dimension $d_c=2$. ``Naive'' power counting suggests that the theory should not
be renormalizable for $d > d_c$, which would mean that the critical behavior 
could no more be inferred from studying the UV limit of the theory, which would
be ill--defined. However, there appears a new fixed point of order $\epsilon$, 
which is IR--{\it unstable}, i.e., UV--{\it stable}, meaning that it governs 
the large--momentum behavior. Hence the theory {\it is} renormalizable despite 
the ``naive'' power--counting arguments. This is precisely what allows for a
description of the critical properties of the model, for the IR--unstable fixed
point physically corresponds to a second--order phase transition in the system.
Indeed, if one inserts the one--loop fixed point value (\ref{3.18}) into 
Eqs.~(\ref{3.19}) and (\ref{3.20}) for the critical exponents at the 
roughening transition, one finds that $z_c = 2 + {\cal O}(\epsilon^3)$ and 
$\chi_c = 0 + {\cal O}(\epsilon^3)$, in accord with the scaling argument by 
Doty and Kosterlitz \cite{dk92}.

\subsection{Two--loop Results}

In this section we return to the field--theoretic method, based on the dynamic
functional (\ref{6}), and present our results from the two--loop calculation.
We shall primarily focus on the analysis and discussion of the results. For 
some details on the calculations and technicalities we refer the reader to the 
Appendices. There is, however, one important point that we have to mention
here. Namely, to two--loop order the situation becomes additionally complicated
with respect to the previous discussion, due to the fact that our massless
field theory produces {\it new} IR singularities at $d = 1$ and $d = 3$ 
(compare Appendix B). These can only be handled by a ``partial'' $\epsilon$
expansion; i.e., we keep all the geometry factors, as explained at length 
above, and expand only in those $d$--dependent factors stemming from the 
integrals as calculated in the dimensional regularization scheme, indicated
with ``$\epsilon$'' instead of ``$d$'' in the Appendix B. This is, we admit, a 
most subtle procedure, and together with the already mentioned features, it
marks the essential difference to the approach by Sun and Plischke 
\cite{sun94}. To our opinion, however, this is the only possible method in 
order to approach the regime $d < 2$ consistently, keeping the exactly known
properties at $d = 1$. For $d > 2$, we shall eventually perform a ``full''
$\epsilon$ expansion in the {\it final} results, where all the problems cited
in the previous subsection have already been accounted for.

>From the two--loop expressions for the singular parts of the two--point vertex
functions (\ref{A21}), (\ref{A22}) in Appendix A, the normalization conditions 
(\ref{3.20})--(\ref{3.22}), and the integrals in Appendix B, we find the 
following results for the renormalization constants
\begin{eqnarray}
   Z_D = 1 &&- {{\hat g_0} \over \epsilon} 
             - (d-1) {{\hat g_0}^2 \over \epsilon}
             - {d-2 \over d} {{\hat g_0}^2 \over 2 \epsilon} \nonumber \\
           &&+ (d-1) {{\hat g_0}^2 \over \epsilon^2}
             + {\cal O}({\hat g_0}^3) \, ,
 \label{3.40} \\
   Z_\nu = 1 &&+ {d-2 \over d} \Biggl[ {{\hat g_0} \over \epsilon}
               + (d-1) {{\hat g_0}^2 \over 2 \epsilon} \nonumber \\
             &&+ {d-2 \over d} {{\hat g_0}^2 \over 2 \epsilon}
               - (d-1) {{\hat g_0}^2 \over 2 \epsilon^2} \Biggr] \nonumber \\
             &&- {d-1 \over d} {{\hat g}_0^2 \over 16 \epsilon} F_\nu(d) 
               + {\cal O}({\hat g_0}^3) \, ,
 \label{3.41}
\end{eqnarray}
where ${\hat g_0} = g_0 C_d \mu^\epsilon$, and we have defined 
\begin{eqnarray}
 F_\nu (d) &&= 4 - 2(6-d) I_{00}(2) + 2d I_{10}(2) \nonumber \\
           && + 21 I_{11}(2)
              - I_{21}(2) - 7 I_{22}(2) + 2 I_{32}(2) \nonumber \\
           && + 4 {\tilde I}_{01}(2) - 4 {\tilde I}_{12}(2) 
    - {12 \over 5 \sqrt{5}} \ln{\sqrt{5}+1 \over \sqrt{5}-1} + {4 \over 5}\, .
 \label{3.42}
\end{eqnarray}
The parameter integrals $I_{rs}(d)$ and ${\tilde I}_{rs}(d)$ are defined in 
Appendix B. The corresponding Wilson flow functions, Eqs.~(\ref{3.30}), are 
given by
\begin{eqnarray}
   \zeta_D = &&- g - {(d-1)(2d-1) \over d} g^2 + {\cal O}(g^3) \, ,
 \label{3.43} \\
   \zeta_\nu = && {d-2 \over d} g + {(d-1)(d-2) \over  d} g^2 \nonumber \\
   &&- {d-1 \over 8 d} g^2 F_\nu (d) + {\cal O}(g^3) \, , 
 \label{3.44} \\
 \beta_g =  g \Biggl( &&d-2 - {2(2d - 3) \over d} g - {(d-1)(5d-7) \over d} g^2
                \nonumber \\
          &&\quad - {d-1 \over 8 d} g^2 F_\nu (d) + {\cal O}(g^3) \Biggr) \, . 
 \label{3.44a}
\end{eqnarray}
(Note that there are no $\epsilon$--dependent terms left in these expressions,
specifically, there are no $1/\epsilon$ poles, which constitutes a very 
non--trivial check to the calculations, along with the fact that at $d=1$ the 
FDT is fulfilled, as required, see Appendix A.)

Most interestingly, the two--loop contributions to these $\zeta$ functions
vanish at $d=1$! That is, there are {\it no singular contributions to the
two--point vertex functions in one dimension}! This is clearly a very valuable
fact for the purpose of justifying a self--consistent approximation as the 
mode--coupling approach, where vertex corrections are neglected \cite{erwin}.
In terms of the fixed points, this means that at $d=1$ the strong--coupling
fixed point in the two--loop approximation is unaltered with respect to the
one--loop result: $g_1^* = 1/2$, and, of course, the critical exponents are
not modified either, $z = 3/2$ and $\chi=1/2$. This is reassuring, for these 
values already follow from a combination of the FDT (thus $\zeta_\nu^* = 
\zeta_D^*$) and the exponent identity (\ref{exponent_identity}), see 
Eqs.~(\ref{3.53}) and (\ref{3.54}) below.

The flow functions (\ref{3.43}) and (\ref{3.44}) permit us to study the 
renormalization--group flow for the renormalized effective coupling constant 
$g$, Eq.~(\ref{3.31}), as a function of the dimensionality of the growth 
problem. The general features of this flow, and the behavior of the ensuing 
{\it three} fixed points $g_0^*$, $g_1^*$ and $g_2^*$ as functions of the 
dimension $d$ are (Fig.~3): 

\par
\noindent (i) Just below the borderline dimension $d_{\rm c}=2$ there are two 
(non--negative) fixed points, the Gaussian fixed point $g_0^* = 0$, which is 
unstable, and one strong coupling fixed point $g_1^*$. The ``weak--coupling''
fixed point $g_0^* = 0$, describing a smooth interface, is IR--unstable. Hence
for $d < 2$ the RG flow always tends to the strong--coupling fixed point 
$g_1^*$, describing a rough surface. (Below $d=1$ there is an additional fixed
point in the physical region, whose value diverges for $d \rightarrow 1$. This
would constitute an unstable fixed point in the flow of the coupling; it 
appears, however, rather doubtful to extrapolate the results of our two--loop
calculations beyond $d=1$, as we had to apply the above--mentioned ``partial''
$\epsilon$ expansion.)

\par
\noindent (ii) At the critical dimension $d_{\rm c}=2$ the strong--coupling 
fixed point $g_1^*$, as obtained from the two--loop calculation, tends to 
infinity. It is not clear, whether this divergence of the strong--coupling 
fixed point at the critical dimension $d_c=2$ is just an artefact of the 
two--loop calculation or a general feature of any finite order in perturbation 
theory. We suppose that it indicates that no {\it finite} strong--coupling
fixed point will emerge to {\it any} order in the perturbation expansion. In 
this sense, there is {\it non--perturbative strong--coupling behavior for all 
$d \geq 2$}. There is still a considerable amount of controversy about the 
existence of an upper critical dimension (not to be confused with the critical 
dimension $d_c=2$ obtained from power counting arguments), at which mean field 
values are recovered. Actually, in view of the analogy to the nonlinear 
$\sigma$ model, the critical dimension $d_c=2$ can be regarded as the {\it 
lower} critical dimension, since for $d_c < 2$ no dynamic roughening transition
takes place [see (i)]. The random directed path on a Cayley tree \cite{cayley1}
represents a possible candidate for a high--dimensional limit. The existence of
a finite upper critical dimension is supported by an expansion by Derrida and 
Cook \cite{cayley2}, who insert finite sections of the high--dimensional 
lattices in place of the nodes of the Cayley tree. The validity of such an 
approach has been criticized by Fisher and Huse \cite{dp2}, however. Also, 
numerical simulations \cite{ft90,wk87,kk89} suggest that there is no upper 
critical dimension. From our two--loop results it is not possible to draw any 
decisive conclusion about the existence of an upper critical dimension. Even 
higher loop orders are most likely not to be very useful for deciding upon the 
question of the existence of an upper critical dimension. What is really needed
is a systematic expansion in $1/d$, or some other controlled expansion capable
of taking into account an infinite set of diagrams. A very recent 
mode--coupling analysis \cite{tu94}, which makes no assumptions about the line 
shape, but is still based on an uncontrolled approximation, indeed yields 
$z < 2$ for all $d$.

\par
\noindent (iii) Above $d_{\rm c}=2$ our two--loop calculations support the
existence of three fixed points. There are two IR stable fixed points, whose 
domain of attraction is separated by a critical fixed point 
$g_2^*=g_{\rm c}^*$, which is IR--unstable (hence UV--stable). The critical 
fixed point $g_{\rm c}^*$ describes a dynamic phase transitions and is 
accessible by perturbation theory (it is of order $\epsilon = d-2$). For 
$g < g_{\rm c}^*$ the RG flow tends towards the weak--coupling fixed point 
$g_{\rm wc}^* = g_0^* = 0$, describing a smooth interface. For effective 
coupling constants larger than $g_{\rm c}^*$ the flow leads to an IR--stable 
strong--coupling fixed point $g_2^* = g_{\rm sc}^*$. This strong--coupling 
fixed point seems to be inaccessible by a perturbational approach. To two--loop
order one thus finds $g_{\rm sc}^* = \infty$. 

This scenario is similar to the results obtained for the $O(n)$ nonlinear 
$\sigma$ model. In a $2+\epsilon$ expansion \cite{nsm} one finds a non--trivial
zero of the corresponding $\beta$ function. This IR--unstable fixed point 
defines the critical temperature, in exactly the same way as the above 
IR--unstable fixed point in the roughening problem defines the critical 
coupling, at which a dynamic phase transition from a smooth to a rough surface 
takes place. This analogy becomes even more apparent if one considers the 
mapping of the Burgers--KPZ equation onto the statistical mechanics of directed
polymers in random media. For $d>2$ Imbrie and Spencer \cite{is88} have shown
rigorously that the polymer undergoes a continuous transition from a low 
temperature pinned phase to a high temperature phase where the disorder is 
irrelevant. The above found critical fixed point controls this transition.

We now proceed to a discussion of the behavior near the different fixed points 
for $d > 2$. In order to relate the values of the $\zeta$ functions at the 
fixed point to the dynamic exponent $z$, and the roughness exponent $\chi$, it 
is convenient to consider the two--point correlation function 
$C_{h h}({\bf q},\omega)$, which acquires the following scaling form
\begin{equation}
  C_{h h} ({\bf q}, \omega) =
  q^{-d-2 \chi -z} {\hat C} ( \omega / q^z ) \, ,
\label{3.45}
\end{equation}
or equivalently (see Section I)
\begin{equation}
  C_{h h} ({\bf x}, t) =
  x^{2 \chi} {\hat C} ( t / x^z ) \, .
\label{3.46}
\end{equation}
We want to analyze how this scaling form and the exponents for the correlation
function are related to the results obtained from renormalized field theory.
Since the bare vertex functions are independent of the arbitrary momentum scale
$\mu$ introduced in the RG procedure, one finds the following RG equation for
the two--point vertex functions $\Gamma (\mu,\nu,D,g,{\bf q}, \omega)$ 
\begin{equation}
 \left[ \mu {\partial \over \partial \mu} + \beta_g {\partial \over \partial g}
       + \zeta_\nu \nu {\partial \over \partial \nu} + 
         \zeta_D D {\partial \over \partial D} \right] \Gamma (.) = 0 \, ,
\label{3.47}
\end{equation}
where we have introduced the abbreviation $(.)=(\mu,\nu,D,g,{\bf q}, \omega)$.
The RG equation is readily solved with the method of characteristics. The 
characteristics $a(l)$ of Eq.~(\ref{3.47}) define running coupling 
``constants'' and parameters into which these transform when $\mu \rightarrow 
\mu(l) = \mu l$. They are given by the solutions to the flow equation of the 
coupling, $l d g / d l = \beta_g(l)$, and the first--order differential 
equations for the parameters $a=D,\nu$
\begin{equation}
   l {d a(l) \over dl} = \zeta_a(l) a(l) \, ,
\label{3.48}
\end{equation}
with the initial conditions $D(l=1) = D$ and $\nu(l=1) = \nu$, namely
\begin{equation}
 a(l) = a \exp \left[ \int_1^l {d \rho \over \rho} \zeta_a (\rho) \right] \, .
\label{3.49}
\end{equation}

Applying a dimensional analysis one finds that $\Gamma_{{\tilde h} {\tilde h}}$
and $\Gamma_{{\tilde h} h}$ have dimensions $D$ and $\nu \mu^2$, respectively.
Hence the solutions of the RG equations read
\begin{eqnarray}
  \Gamma_{{\tilde h} h} (.) =
  &&\mu^2 l^2 \nu(l) \Gamma_{{\tilde h} h} 
                        \left( 
                              {{\bf q} \over \mu l},
                              {\omega \over \mu^2 l^2 \nu(l)}, g(l)
                        \right) \, , \label{3.50} \\
  \Gamma_{{\tilde h} {\tilde h} } (.) =
  &&D(l) \Gamma_{{\tilde h} {\tilde h} } 
                        \left( 
                              {{\bf q} \over \mu l},
                              {\omega \over \mu^2 l^2 \nu(l)}, g(l)
                        \right) \, . \label{3.51} 
\end{eqnarray}
Since the two--point correlation function is related to the two--point vertex
function by $C = {\Gamma_{{\tilde h} {\tilde h}} / |\Gamma_{{\tilde h} h}|^2}$,
one gets at a fixed point, with the matching condition ${\bf q} / \mu l = 1$,
\begin{equation}
    C (\mu,\nu,D,g,{\bf q}, \omega) = q^{-4-2\zeta_\nu^*+\zeta_D^*}
    {\hat C} \left({\omega \over q^{2+\zeta_\nu^*}} \right)  \, .
\label{3.52}
\end{equation}
Hence we arrive at the following, already mentioned relations
\begin{eqnarray}
  \chi = &&1 - {d \over 2} + {\zeta_\nu^* - \zeta_D^* \over 2} \, ,
  \label{3.53} \\
  z = &&2 + \zeta_\nu^* \, . \label{3.54}
\end{eqnarray}

At the weak--coupling fixed point, $g_0^*=0$, we have $\chi = 1 - d / 2$
and $z=2$. At any non--zero fixed point $g^* \neq 0$ one gets 
\begin{equation}
  d-2+\zeta_D^* - 3\zeta_\nu^* =0
\label{3.55}
\end{equation}
from $\beta_g (g^*) = 0$. Note that this exponent identity results ultimately 
from the Galilean invariance of the Burgers--KPZ equation. Hence, we find
\begin{eqnarray}
  \chi = && - \zeta_\nu^* \, ,
  \label{3.56} \\
  z = &&2 + \zeta_\nu^* \, .
  \label{3.57}
\end{eqnarray}
>From these relations one can easily infer the exponent identity $\chi+z=2$
already mentioned in Section I.

Let us now investigate the two--loop results at the IR--unstable (critical) 
fixed point $g_{\rm c}^*$. One can show that $F_\nu(d) = 8 + {\cal O}
(\epsilon)$, where the ${\cal O}(\epsilon)$ coeffficient can also be determined
numerically $F_\nu(d) \approx 8 - 4.0797 \epsilon$ [see Eq.~(\ref{B13})]. 
Hence, one finds that in a consistent expansion in both $g$ and $\epsilon$
\begin{eqnarray}
  \zeta_D = &&-g -{3 \over 2} g^2 + 
   {\cal O}(g^3,g^2\epsilon,g\epsilon^2,\epsilon^3) \, ,
  \label{3.58}  \\
  \zeta_\nu = &&{\epsilon \over 2} g -{1 \over 2} g^2 + 
  {\cal O}(g^3,g^2\epsilon,g\epsilon^2,\epsilon^3) 
  \label{3.59} 
\end{eqnarray}
where $\epsilon = d-2$. With Eq.~(\ref{3.31}) this yields for the $\beta$
function
\begin{equation}
  \beta_g = g \left( \epsilon - {2+3\epsilon \over 2} g + 
              {\cal O}(g^3,g^2\epsilon,g\epsilon^2,\epsilon^3) \right) \, ,
\label{3.60}
\end{equation}
which is identical to the one--loop function since all the ${\cal O}(g^2)$ 
corrections cancel! The resulting critical fixed point is $g_{\rm c}^* = 
\epsilon - 3 \epsilon^2 / 2 + {\cal O}(\epsilon^3)$, and there is {\it no
finite strong--coupling fixed point}. Inserting this result for the unstable 
fixed point into the $\zeta$ functions one realizes that {\it all} 
${\cal O}(\epsilon^2)$ {\it corrections cancel}, i.e.,
\begin{eqnarray}
  \zeta_D^* = && - \epsilon + {\cal O} ( \epsilon^3) \, ,
  \label{3.61}  \\
  \zeta_\nu^* = &&0 + {\cal O} ( \epsilon^3) \, .
  \label{3.62} 
\end{eqnarray}
Hence our two--loop calculations confirm the results by Doty and Kosterlitz 
\cite{dk92}, that $z_c = 2$ and $\chi_c = 0$ at the roughening transition. Note
that in our perturbation expansion this result is due to a most remarkable and 
not at all obvious cancellation of very different contributions. This is a 
very reassuring feature of our method again, at least if one follows the 
considerations in Ref.~\cite{dk92}.

Now let us reverse the argument, and assume that $z_c=2$ is an {\it exact} 
result at the critical fixed point. Then, with Eqs.~(\ref{3.55}) and 
(\ref{3.57}) one gets at the weak--coupling fixed point 
\begin{eqnarray}
  (d-2) + \zeta_D (g_{\rm c}^*) = && 0 \, ,
  \label{3.63}  \\
  \zeta_\nu (g_{\rm c}^*)  = &&0 \, .
  \label{3.64} 
\end{eqnarray}
Therefore, it should also be possible to determine the value of the critical
fixed point from one of the latter equations. In fact, one obtains from 
Eq.~(\ref{3.63}) the same critical fixed point as from the zero of the
$\beta$ function. However, in order to determine the value of $g_c^*$ up to 
terms of order ${\cal O}(\epsilon^3)$ from Eq.~(\ref{3.64}) one needs the 
coefficient of the ${\cal O}(g^2)$ term and the ${\cal O}(g^3)$ term of 
$\zeta_\nu$ up to terms of order ${\cal O}(\epsilon^2)$ and 
${\cal O}(\epsilon^1)$, respectively. For the calculation of the fixed point 
value from the zeros of the $\beta$ function, we had to know the 
${\cal O}(g^2)$ term of $\zeta_\nu$ up to terms of order ${\cal O}(\epsilon)$, 
only. From Eq.~(\ref{3.44}) one obtains upon including an unknown three--loop 
term
\begin{equation}
  \zeta_\nu = {g \over d} 
              \left[ 
              \epsilon + 
              \left(-1+{4.0797 \over 8}\epsilon \right) g + 
              A g^2 \right] \, ,
  \label{3.65}
\end{equation}
where A is a constant of order ${\cal O}(1)$. Eq.~(\ref{3.64}) yields for the 
critical fixed point: $g^*_{\rm c} = \epsilon - 3 \epsilon^2 / 2$, only if
$A=-\left( 3/2 + 4.0797/8 \right)$, i.e., one can determine the 
three--loop correction for the renormalization of the surface tension $\nu$.

Let us comment on the $d$ dependence of the critical fixed point. Keeping the
full dependence of the $\zeta$ function for the noise amplitude on $d$, one
obtains from Eqs.~(\ref{3.43}) and (\ref{3.63}) for the critical fixed point
\begin{equation}
   g^*_{\rm c} = {-1+\sqrt{1+4 (d-1) (2d-1) (d-2) / d} 
                   \over 2(d-1) (2d-1) / d} \, ,
\label{3.66}
\end{equation}
which in the limit of large $d$ reduces to a finite value $g_{\rm c}^* 
\rightarrow 1 / \sqrt{2}$. It would be interesting to see how this $d$
dependence of the fixed point compares with numerical simulations. Note, 
however, that the scale of $g_c^*$ is non--universal and depends on the precise
definition of the coupling $g$.

We close this section by a discussion of the RG equations for $d > 2$, and its
implications on the kinetic roughening transition described by the IR--unstable
(repulsive) fixed point $g_c^*$. We remark that a detailed analysis of the 
crossover scaling behavior in the weak--coupling regime at the one--loop level 
has been given by Nattermann and Tang \cite{nat92}. We have shown above that 
the exponents at the transition are given by $z_c = 2 + {\cal O} (\epsilon^3)$ 
and $\chi_c = 0 + {\cal O} (\epsilon^3)$. In the smooth phase (i.e., for 
coupling constants $g<g_c^*$) the effective coupling approaches zero. 
Therefore, the long--distance and long--time properties in this phase may be 
obtained from a perturbative RG study. (In the rough phase a perturbative 
expansion is not possible due to the strong--coupling behavior.)

The proximity to the critical point can be measured in terms of the control 
parameter $\delta = (g_c^*-g)/g_c^*$, and, quite analogously to the treatment
of the nonlinear $\sigma$ model at its critical point \cite{nsm}, we define 
a correlation length $\xi(g)$ via the solution of the differential equation
\begin{equation}
  \beta_g(g) {d \xi(g) \over dg} = \xi(g) \, .
\label{3.67}
\end{equation}
Since $\xi$ has the dimension of a length this can also be written as
\begin{equation}
    \left[ \mu {\partial \over \partial \mu} + \beta_g(g) 
           {\partial \over \partial g} \right] \xi(g,\mu) = 0 \, ,
\label{3.67a}
\end{equation}
i.e.,
\begin{equation}
\xi(g,\mu) = \mu^{-1} g^{1/(d-2)} \exp
 \left[ \int_0^g dg^\prime \left( {1 \over \beta_g(g^\prime)} - 
       { 1 \over (d-2) g^\prime } \right) \right]
\label{3.68}
\end{equation}
Close to the critical fixed point we can write
\begin{equation}
  {d \xi \over d g} = {\xi \over \beta_g(g(l))}
                    = {\xi \over [g(l)-g_c^*] \beta_g^\prime(g_c^*)} \, ,
\label{3.69}
\end{equation}
the solution of which is given by
\begin{equation}
  \xi (g) \propto |g-g_c^*|^{1/\beta_g^\prime (g_c^*)} 
  \propto |\delta|^{-\nu} \, ,
\label{3.70}
\end{equation}
where $\beta_g^\prime (g_c^*) = d \beta_g (g) / dg |_{g_c^*}$. Hence the 
critical exponent for the correlation length is determined by the first 
derivative of the $\beta$ function at the critical fixed point 
$\nu = -1/\beta_g^\prime (g_c^*)$. Since we have shown above that all two--loop 
contributions to this $\beta$ function cancel, we find that the correlation 
length exponent is
\begin{equation}
  1/\nu = \epsilon + {\cal O}(\epsilon^3) \, .
\label{3.70a}
\end{equation}
To one--loop order, this result was already obtained by Nattermann and Tang 
\cite{nat92}, who also discuss how the above scaling is modified to a 
logarithmic form at $d_c=2$.

Let us now solve the RG equations, Eq.~(\ref{3.47}), in such a way that the 
scaling behavior at the dynamic phase transition becomes apparent (see also
Ref.~\cite{nat92}). Dimensional analysis tells us that
\begin{eqnarray}
  \Gamma_{{\tilde h}{\tilde h}} (\mu,\nu,D,g,{\bf q}, \omega) = 
  && D {\hat \Gamma}_{{\tilde h}{\tilde h}}
     \left( {q \over \mu}, {\omega \over \mu^2 \nu} \right) \, ,
  \label{3.71a} \\
  \Gamma_{{\tilde h}h} (\mu,\nu,D,g,{\bf q}, \omega) = 
  && \mu^2 \nu {\hat \Gamma}_{{\tilde h}h}
     \left( {q \over \mu}, {\omega \over \mu^2 \nu} \right) \, .
\label{3.71b}
\end{eqnarray}
Thus the solution of the RG equation can be written as
\begin{eqnarray}
\Gamma_{{\tilde h}{\tilde h}} (.) = 
&&D (g)
\gamma_{{\tilde h}{\tilde h}} 
       \left( q \xi(g), {\omega \xi^2(g) \over \nu (g)}
       \right) \, , \label{3.72a} \\
\Gamma_{{\tilde h}h} (.) =
&&\xi(g)^{-2} \nu (g)
\gamma_{{\tilde h}h} 
       \left( q \xi(g), {\omega \xi^2(g) \over \nu (g)}
       \right) \, , \label{3.72b} 
\end{eqnarray}
with 
\begin{eqnarray}
D(g) &&= D \exp \left\{
        - \int_0^g {dg^\prime \over \beta(g^\prime)} \zeta_D(g^\prime)
        \right\} \, , \label{3.73a} \\
\nu(g) &&=  \nu 
            \exp 
            \left\{ - \int_0^g {dg^\prime \over \beta(g^\prime)} 
                      \zeta_\nu (g^\prime)
            \right\} \, . \label{3.73b} 
\end{eqnarray}
In summary, one obtains the the following dynamic scaling behavior of 
the correlation function
\begin{equation}
  C(\delta, {\bf q}, \omega) =
  \, M(\xi(g)) \, \xi^4(g) \, {\hat C} 
  \left( {q\xi(g)}, {\omega \over \omega_c (\xi(g))} \right) \, , 
\label{3.71}
\end{equation}
where we have defined the quantity
\begin{equation}
  M(\xi(g)) = \exp \left\{-
                        \int_{0}^g {dg \over \beta(g)} 
                        \left[ \zeta_D(g) - 2 \zeta_\nu (g) \right] 
                        \right\} \, .
\label{3.72}
\end{equation}
We have also introduced a characteristic frequency by
\begin{equation}
  \omega_c(\xi(g)) = \mu^2 \xi^{-2} 
                     \exp \left\{- 
                          \int_{0}^g {dg \over \beta(g)} \, \zeta_\nu (g)
                          \right\} \, .
\label{3.73}
\end{equation}
Close to the critical fixed point the latter equation reduces to 
$\omega_c(\xi(g)) \propto \xi^{-z_c}$, with the dynamical critical exponent
$z_c = 2 + \zeta_\nu^*$. One should note that this dynamic scaling form is 
quite analogous to the corresponding scaling in the nonlinear $\sigma$ model 
\cite{nsm}, and the same conclusions drawn there apply for the roughening 
transition as well. Let us contrast the above scaling behavior with the 
well--known scaling behavior of a standard $\phi^4$ model close to its critical 
point. The $\phi^4$ model depends on two coupling constants, the coefficient of 
the $\phi^2$ term which plays the role of the control parameter (reduced 
temperature) corresponding to the reduced effective coupling $\delta$ defined 
above, and the coefficient of the $\phi^4$ term which has {\it no} equivalent 
here. The scaling form of the $\phi^4$ model is obtained at the {\it 
IR--stable} fixed point, e.g. obtained by an $\epsilon$ expansion near the 
upper critical dimension $4$. In the present case, however, there is not only 
scaling behavior at the critical fixed point, but in the entire ``smooth'' 
phase. This behavior, Eq.~(\ref{3.71}), is most similar to that of $O(n)$
symmetric models in the low temperature phase, and more closely resembles the 
form known from crossover scaling behavior \cite{nat92}. In order to study the
crossover scaling behavior in the ``weak coupling'' phase, we use the following
``resummed'' expressions for the $\zeta$ and $\beta$ functions
\begin{eqnarray}
  \zeta_D = &&-{ g \over 1 - 3/2 g} \, ,
  \label{3.74}  \\
  \zeta_\nu = &&{g \over d} \left( \epsilon -{g \over 1 - 3/2 g} \right) 
    \, ,  \label{3.75}  \\
  \beta_g = &&g \left( \epsilon - {2+3\epsilon \over 2} g \right) \, ,
\label{3.76}
\end{eqnarray}
which are identical to the expressions above, but also take into account that
the fixed point values can be equally well obtained from the zero of the
$\beta$ function and Eqs.~(\ref{3.63}) and (\ref{3.64}).

First, we consider the crossover of the correlation length as a function of the
effective coupling $g$. As depicted in Fig.~4, the correlation length crosses 
over from $\xi \propto g^{1/\epsilon}$ at small couplings to $\xi \propto 
|g-g_c^*|^{-\nu}$ as $g$ approaches the critical coupling $g_c^*$. The 
crossover function shown in Fig.~4 is universal, i.e. all the non--universal 
scales may be absorbed in an amplitude $\xi_0$ for the correlation length. 
Specifically, the location of the crossover $g_{\rm cross}$ can be obtained 
from Fig.~4. Similar crossover behavior is found for the noise amplitude $D$ 
and the surface tension $\nu$, as shown in Fig.~5. The asymptotic behavior at 
the critical fixed point is given by
\begin{eqnarray}
  D (g) \propto &&|g-g_c^*|^{-\epsilon \nu} = |g-g_c^*|^{-1} \, ,
  \label{3.77} \\
  \nu (g) \propto &&|g-g_c^*|^{0} \, .
  \label{3.78}
\end{eqnarray}
The noise amplitude shows a crossover from an exponential increase $D(g) 
\propto e^{g/\epsilon}$ at small values of $g$ to a power law divergence $D(g) 
\propto |g-g_c^*|^{- \epsilon \nu}$ as $g$ approaches the critical coupling 
$g_c^*$. The renormalized surface tension $\nu(g)$ crosses over from 
exponentially decreasing $\nu (g) \propto e^{-g/2}$ to a constant at 
criticality. Finally, the crossover behavior of the characteristic frequency is
depicted in Fig.~6. It crosses over from $\omega_c (g) \propto g^{-2/\epsilon} 
e^{-g/2}$ at small couplings to $\omega_c (g) \propto |g-g_c^*|^{-2}$ as $g$ 
approaches the critical coupling $g_c^*$. 

It would be interesting to test the validity of the above crossover scaling,
including the from of the crossover functions, by numerical simulation.

\section{Summary and Conclusions}

In this paper we have given a systematic analysis of the Burgers--KPZ equation 
in $d+1$ dimensions by dynamic renormalization group theory. We have paid 
special attention to the interconnections of the various renormalization group 
techniques.

Let us briefly summarize our main conclusions. For the roughening dynamics as
described by the Burgers--KPZ equation we find the following results:
(1) The roughening transition of the Burgers--KPZ equation is understood in 
terms of an IR--unstable fixed point above the critical dimension $d_c=2$. The 
value of this critical fixed point is accessible by a $2+\epsilon$ expansion, 
similar to an analogous expansion for the nonlinear $\sigma$ model. The 
critical properties of this transition are characterized by one independent
exponent. Doty and Kosterlitz have argued on the basis of a standard scaling 
argument that the dynamic exponent $z_c$ at the dynamic roughening transition 
should be exactly equal to $2$. From our two--loop calculation we find that 
$z_c = 2 + {\cal O} (\epsilon^3)$, which supports their considerations.
We have further analyzed the scaling behavior near the transition and in the
smooth phase. Introducing a control parameter $\delta = (g_c^* - g)/ g_c^*$,
which measures the distance to the critical point, one can introduce a 
correlation length $\xi$ and calculate the corresponding exponent $\nu$. We 
find that $\nu = {1 / \epsilon} + {\cal O} (\epsilon^3)$.
(2) Below the borderline dimension $d_c=2$ there appear two fixed points, an 
IR--unstable Gaussian fixed point and an IR--stable ``strong--coupling'' fixed
point. Note that the strong--coupling fixed point is {\it not} of order 
${\cal O}(\epsilon)$ and there is apparently also no other known small 
parameter. Hence, the phase described by this strong--coupling fixed point is 
not accessible by a controlled perturbation theory, and one needs 
non--perturbative methods (e.g. mode--coupling theories) to access the physics 
of the rough phase. The ``weak--coupling'' fixed point describes a smooth 
interface. But, since this fixed point is always unstable, the interface is 
always rough below $d_c=2$.
(3) At the critical dimension the strong--coupling fixed point, as obtained 
from the two--loop calculation, tends to infinity. It is not clear, whether 
this divergence of the strong--coupling fixed point is just an artefact of the 
two--loop calculation or a general feature of any finite order in perturbation
theory. We suppose that it indicates that there is non--perturbative 
strong--coupling behavior for {\it all} $d \geq 2$. 

We have also highlighted the importance of carefully considering the structure 
of the dimensional regularization method. As explained in detail in Section 
III, some care is required in using this method, especially in distinguishing 
between the dimensional dependence contained in $1/\epsilon$ poles and 
$d$--dependent factors originating from purely geometric properties of the loop
integrals. Whereas the $1/\epsilon$ poles constitute essentially nothing but a 
quite convenient way of keeping track of the UV divergences in the perturbation
expansion, all other $d$--dependent factors characterize the symmetry and 
internal structure of the model. Note that with these precautions the results 
for the renormalization factors quite naturally obey the FDT in $d=1$ 
dimension; it is not necessary to perform a separate treatment of this case.

The two--loop calculations allow for no decisive conclusion about the existence
of an upper critical dimension, at which mean field values for the exponents 
are recovered. To decide upon the existence of an upper critical dimension it 
would be highly useful to do a systematic expansion in $1/d$ within the 
framework of the Burgers--KPZ equation. To our knowledge, such an expansion is 
not yet available at the present time.

Very recent results by Tu \cite{tu94}, who has solved the mode--coupling 
equations without assuming any specific line shape, indicate that $z<2$ for all
dimensions $d$. However, the results obtained from mode--coupling theories are 
not quite conclusive, since it is not known in what sense such theories 
constitute a controlled expansion. Nevertheless, resummation techniques such as
the mode--coupling theory are at present the only available analytical 
technique to study strong--coupling behavior. 

For the ($1+1$)--dimensional case there is very close agreement between 
numerical simulations and the results obtained from mode--coupling theory 
\cite{erwin}. Our two--loop calculations demonstrate that for $d=1$ {\it all} 
two--loop corrections vanish. This is a highly non--trivial result, because for
this to be the case, the vertex corrections entering the calculations of the 
correlation and response functions must cancel each other. It is exactly those 
vertex corrections, which are neglected by mode-coupling theory, however. Hence
our RG analysis provides a distinct hint why this approach has been so 
successful in $1+1$ dimension. Yet, further analysis is required to clarify 
this point \cite{fht94}.

Summarizing, it would thus be of interest to further investigate the behavior 
of the strong--coupling fixed point by a systematic $1/d$ expansion, as well 
as by mode--coupling theory in high dimensions. In addition, it would be 
interesting to know how the value of the critical fixed point (obtained within
the two--loop calculation) --- especially its $d$ dependence --- and the 
crossover scaling analysis in the smooth phase compare with results from 
numerical simulations.

\vskip 1truecm

\acknowledgements{It is a pleasure to acknowledge helpful discussions with
T.~Hwa, D.S.~Fisher, M.~Kardar, T.~Nattermann, G.~Zumbach, L.~Balents, and 
T.~Sun. The work of E.F. and U.C.T has been supported by the Deutsche 
Forschungsgemeinschaft (DFG) under Contracts No. Fr. 850/2-1,2 and Ta. 177/1-1,
respectively.}

\vfill 

\eject

\newpage

\onecolumn

\widetext

\section{Appendix}

In the appendices, we list the explicit results for the perturbation theory to
two--loop order. Appendix A comprises the Feynman diagrams and the 
corresponding analytical expressions, while we present the evaluation of the 
relevant integrals in dimensional regularization in Appendix B. 

\subsection{Two--loop Perturbation Theory for the Two--point Vertex Functions}

This appendix comprises the Feynman diagrams to two--loop order for the 
Burgers -- Kardar-Parisi-Zhang equation, and the corresponding momentum 
integrals. The integrations over the internal frequencies have already been 
performed using the residue theorem.

We start with a list of the contributions to two--loop order to the fully 
wavevector-- and frequency--dependent two--point vertex functions. In writing
down the diagrammatic expansion for the dynamic functional one has to take
into account restrictions which follow from causality.
In section II we have not written down the Jacobian ${\cal J} [h] = {\cal D} 
[\eta] / {\cal D} [h]$. In general the Jacobian depends on the discretization 
of the Langevin equation (needed to give a proper definition to the path 
integral). As can be shown quite generally~\cite{bjw76} the Jacobian cancels 
the equal time contractions of the field $h$ and the response field 
${\tilde h}$. Keeping in mind (or by choosing a discretization with the 
Jacobian equal to $1$) one can omit the Jacobian in the dynamic functional. The
Feynman diagrams, which account for the restrictions imposed by 
causality~\cite{fnote4}, are depicted in Fig.~7 and Fig.~8 for 
$\Gamma_{{\tilde h} {\tilde h}}({\bf q},\omega)$ and 
$\Gamma_{{\tilde h} h}({\bf q},\omega)$, respectively. In case of the former,
adding the terms corresponding to both opposite time directions, respectively,
provides some simplifications. The corresponding analytical expressions are:
\begin{eqnarray}
   &&\Gamma_{{\tilde h} {\tilde h}}({\bf q},\omega) \, :  \nonumber \\
   &&(a) + (b) = - 2 D_0 - {\lambda_0^2 D_0^2 \over 2 \nu_0^3} {\rm Re} 
      \int_p {(q^2/4-p^2)^2 \over ({\bf q}/2+{\bf p})^2 ({\bf q}/2-{\bf p})^2} 
             {1 \over i\omega/2\nu_0+q^2/4+p^2}                           \, ,
 \label{A1} \\
   &&(c) + (d) = {\lambda_0^4 D_0^3 \over 8 \nu_0^6} {\rm Re}
    \int_p {(q^2/4-p^2)^2 \over ({\bf q}/2+{\bf p})^2 ({\bf q}/2-{\bf p})^4}
    \int_k {({\bf q}/4-{\bf p}/2)^2-k^2 \over i\omega/2\nu_0+q^2/4+p^2} \times
 \label{A2} \\    
   &&\qquad \qquad \qquad \times 
    {({\bf q}/2-{\bf p})^2 [({\bf q}/4-{\bf p}/2)^2+k^2]
                              -2[({\bf q}{\bf k})/2-({\bf p}{\bf k})]^2
     \over ({\bf q}/4-{\bf p}/2+{\bf k})^2 ({\bf q}/4-{\bf p}/2-{\bf k})^2
                           [3({\bf q}/2-{\bf p})^2/4+k^2]} \times \nonumber \\
   &&\qquad \times \biggl[ 2 + {({\bf q}/2-{\bf p})^2 \over 
         i\omega/2\nu_0+({\bf q}/2+{\bf p})^2/2+({\bf q}/2-{\bf p})^2/4+k^2}
     \biggl( 1 + {3({\bf q}/2-{\bf p})^2/4+k^2 \over i\omega/2\nu_0+q^2/4+p^2}
     \biggr) \biggr] \, , \nonumber \\
   &&(e) = - {\lambda_0^4 D_0^3 \over 8 \nu_0^6} {\rm Re}
     \int_p {(q^2/4-p^2)^2 \over ({\bf q}/2+{\bf p})^2 ({\bf q}/2-{\bf p})^2} 
            {1 \over i\omega/2\nu_0+q^2/4+p^2} \times
 \label{A3} \\
   &&\qquad \qquad \qquad \qquad \int_k {[({\bf q}/4-{\bf p}/2)^2-k^2]^2
     \over ({\bf q}/4-{\bf p}/2+{\bf k})^2 ({\bf q}/4-{\bf p}/2-{\bf k})^2
                           [3({\bf q}/2-{\bf p})^2/4+k^2]} \times \nonumber \\
   &&\qquad \qquad \qquad \qquad \qquad \qquad \times
     \biggl( 2 + {({\bf q}/2-{\bf p})^2 \over 
       i\omega/2\nu_0+({\bf q}/2+{\bf p})^2/2+({\bf q}/2-{\bf p})^2/4+k^2}     
     \biggr)                                                 \, , \nonumber \\
   &&(f) + (g) = - {\lambda_0^4 D_0^3 \over \nu_0^6} {\rm Re}
     \int_p {q^2/4-p^2 \over ({\bf q}/2+{\bf p})^2 ({\bf q}/2-{\bf p})^2}
     \int_k {({\bf q}{\bf k})/2+({\bf p}{\bf k}) \over 
                           ({\bf q}/2-{\bf p}-{\bf k})^2} \times 
 \label{A4} \\                                   &&\qquad \qquad \qquad \times 
     {q^2/4-({\bf p}+{\bf k})^2 \over i\omega/2\nu_0+q^2/4+({\bf p}+{\bf k})^2}
     {({\bf q}{\bf k})/2-({\bf p}{\bf k})-({\bf q}/2-{\bf p})^2 \over 
                       ({\bf q}/2-{\bf p}-{\bf k})^2+({\bf q}/2-{\bf p})^2+k^2}
     \times                                                       \nonumber \\ 
   &&\times \biggl( {\rm Re} {1 \over i\omega/2\nu_0+q^2/4+p^2} + 
            {1 \over i\omega/2\nu_0+q^2/4+p^2}{({\bf q}/2-{\bf p})^2 \over
         i\omega/\nu_0+({\bf q}/2-{\bf p}-{\bf k})^2+({\bf q}/2+{\bf p})^2+k^2}
     \biggr) \, ,                                                 \nonumber \\
   &&(h) + (i) = {\lambda_0^4 D_0^3 \over \nu_0^6} {\rm Re}
    \int_p {q^2/4-p^2 \over ({\bf q}/2+{\bf p})^2 ({\bf q}/2-{\bf p})^2} \int_k
      {({\bf q}{\bf k})^2/4-({\bf p}{\bf k})^2 \over k^2} \times 
 \label{A5} \\                                   &&\qquad \qquad \qquad \times 
     {q^2/4-({\bf p}+{\bf k})^2 \over i\omega/2\nu_0+q^2/4+({\bf p}+{\bf k})^2}
     {1 \over ({\bf q}/2-{\bf p}-{\bf k})^2+({\bf q}/2-{\bf p})^2+k^2}
     \times                                                       \nonumber \\ 
   &&\times \biggl( {\rm Re} {1 \over i\omega/2\nu_0+q^2/4+p^2} + 
            {1 \over i\omega/2\nu_0+q^2/4+p^2}{({\bf q}/2-{\bf p})^2 \over
         i\omega/\nu_0+({\bf q}/2-{\bf p}-{\bf k})^2+({\bf q}/2+{\bf p})^2+k^2}
     \biggr) \, ,                                                 \nonumber \\
   &&(k) = - {\lambda_0^4 D_0^3 \over 2 \nu_0^6} {\rm Re}
     \int_p {q^2/4-p^2 \over ({\bf q}/2+{\bf p})^2}
     \int_k {({\bf q}{\bf k})/2+({\bf p}{\bf k}) \over 
                       k^2 ({\bf q}/2-{\bf p}-{\bf k})^2} \times 
 \label{A6} \\                                   &&\qquad \qquad \qquad \times 
     {q^2/4-({\bf p}+{\bf k})^2 \over i\omega/2\nu_0+q^2/4+({\bf p}+{\bf k})^2}
     {({\bf q}{\bf k})/2-({\bf p}{\bf k})-k^2 \over 
                       ({\bf q}/2-{\bf p}-{\bf k})^2+({\bf q}/2-{\bf p})^2+k^2}
     \times                                                       \nonumber \\ 
   &&\qquad \qquad \qquad \times \biggl( {1 \over -i\omega/2\nu_0+q^2/4+p^2} + 
{2 \over i\omega/\nu_0+({\bf q}/2-{\bf p}-{\bf k})^2+({\bf q}/2+{\bf p})^2+k^2}
     \biggr) \, ;                                                 \nonumber
\end{eqnarray}
\begin{eqnarray}
   &&\Gamma_{{\tilde h} h}({\bf q},\omega) \, :                \nonumber \\
   &&(a) + (b) = i \omega + \nu_0 q^2 + {\lambda_0^2 D_0 \over 4 \nu_0^2} 
      \int_p {q^2/4-p^2 \over ({\bf q}/2-{\bf p})^2}
             {q^2-2({\bf q}{\bf p}) \over i\omega/2\nu_0+q^2/4+p^2}     \, ,
 \label{A7} \\
   &&(c) = - {\lambda_0^4 D_0^2 \over 32 \nu_0^5}
    \int_p {q^2/4-p^2 \over ({\bf q}/2+{\bf p})^2}
           {q^2+2({\bf q}{\bf p}) \over (i\omega/2\nu_0+q^2/4+p^2)^2} 
    \int_k {({\bf q}/4-{\bf p}/2)^2-k^2 \over
        ({\bf q}/4-{\bf p}/2+{\bf k})^2 ({\bf q}/4-{\bf p}/2-{\bf k})^2} \times
 \nonumber \\                       &&\qquad \qquad \qquad \qquad \qquad \times
    {({\bf q}/2-{\bf p})^2 [({\bf q}/4-{\bf p}/2)^2+k^2]
                           -2[({\bf q}{\bf k})/2-({\bf p}{\bf k})]^2 \over
         i\omega/2\nu_0+({\bf q}/2+{\bf p})^2/2+({\bf q}/2-{\bf p})^2/4+k^2}
 \, , \label{A8} \\
   &&(d) = - {\lambda_0^4 D_0^2 \over 32 \nu_0^5}
    \int_p {q^2/4-p^2 \over ({\bf q}/2-{\bf p})^4}
           {q^2-2({\bf q}{\bf p}) \over i\omega/2\nu_0+q^2/4+p^2} \times
 \label{A9} \\    
   &&\qquad \times \int_k {({\bf q}/4-{\bf p}/2)^2-k^2 \over 
                                              3({\bf q}/2-{\bf p})^2/4+k^2} 
           {({\bf q}/2-{\bf p})^2 [({\bf q}/4-{\bf p}/2)^2+k^2]
                           -2[({\bf q}{\bf k})/2-({\bf p}{\bf k})]^2 \over
            ({\bf q}/4-{\bf p}/2+{\bf k})^2 ({\bf q}/4-{\bf p}/2-{\bf k})^2}
 \times \nonumber \\ &&\qquad \times \biggl[ 1 + {({\bf q}/2-{\bf p})^2 \over 
         i\omega/2\nu_0+({\bf q}/2+{\bf p})^2/2+({\bf q}/2-{\bf p})^2/4+k^2}
     \biggl( 1 + {3({\bf q}/2-{\bf p})^2/4+k^2 \over i\omega/2\nu_0+q^2/4+p^2}
       \biggr) \biggr]                                       \, , \nonumber \\
   &&(e) = - {\lambda_0^4 D_0^2 \over 32 \nu_0^5}
    \int_p {q^2/4-p^2 \over ({\bf q}/2-{\bf p})^4}
           {q^2-2({\bf q}{\bf p}) \over i\omega/2\nu_0+q^2/4+p^2} \times
 \label{A10} \\    
   &&\qquad \qquad \times \int_k {({\bf q}/4-{\bf p}/2)^2-k^2 \over 
                                              3({\bf q}/2-{\bf p})^2/4+k^2} 
           {({\bf q}/2-{\bf p})^2 [({\bf q}/4-{\bf p}/2)^2+k^2]
                           -2[({\bf q}{\bf k})/2-({\bf p}{\bf k})]^2 \over
           ({\bf q}/4-{\bf p}/2+{\bf k})^2 ({\bf q}/4-{\bf p}/2-{\bf k})^2} 
                                                             \, , \nonumber \\
   &&(f) = {\lambda_0^4 D_0^2 \over 32 \nu_0^5} \int_p {q^2/4-p^2 \over 
 ({\bf q}/2-{\bf p})^2} {q^2-2({\bf q}{\bf p}) \over i\omega/2\nu_0+q^2/4+p^2} 
    \int_k {1 \over 3({\bf q}/2-{\bf p})^2/4+k^2} \times
 \label{A11} \\    
   &&\quad \times {[({\bf q}/4-{\bf p}/2)^2-k^2]^2 \over 
            ({\bf q}/4-{\bf p}/2+{\bf k})^2 ({\bf q}/4-{\bf p}/2-{\bf k})^2}
           \biggl( 2 + {({\bf q}/2-{\bf p})^2 \over 
   i\omega/2\nu_0+({\bf q}/2+{\bf p})^2/2+({\bf q}/2-{\bf p})^2/4+k^2} \biggr) 
                                                             \, , \nonumber \\
   &&(g) = {\lambda_0^4 D_0^2 \over 8 \nu_0^5} \int_p {q^2-2({\bf q}{\bf p}) 
                       \over ({\bf q}/2-{\bf p})^2 (i\omega/2\nu_0+q^2/4+p^2)} 
    \int_k {({\bf q}{\bf k})/2+({\bf p}{\bf k}) \over 
                                         ({\bf q}/2-{\bf p}-{\bf k})^2} \times 
 \label{A12} \\                                  &&\qquad \qquad \qquad \times 
    {q^2/4-({\bf p}+{\bf k})^2 \over i\omega/2\nu_0+q^2/4+({\bf p}+{\bf k})^2}
     {({\bf q}{\bf k})/2-({\bf p}{\bf k})-({\bf q}/2-{\bf p})^2 \over 
                      ({\bf q}/2-{\bf p}-{\bf k})^2+({\bf q}/2-{\bf p})^2+k^2}
     \times                                                       \nonumber \\ 
   &&\qquad \qquad \qquad \qquad\times \biggl( 1 + {2 ({\bf q}/2-{\bf p})^2 
 \over  i\omega/\nu_0+({\bf q}/2-{\bf p}-{\bf k})^2+({\bf q}/2+{\bf p})^2+k^2} 
     \biggr)                                                 \, , \nonumber \\
   &&(h) = {\lambda_0^4 D_0^2 \over 8 \nu_0^5} \int_p {q^2+2({\bf q}{\bf p}) 
                       \over ({\bf q}/2+{\bf p})^2 (i\omega/2\nu_0+q^2/4+p^2)} 
    \int_k {({\bf q}{\bf k})/2+({\bf p}{\bf k}) \over 
                                         ({\bf q}/2-{\bf p}-{\bf k})^2} \times 
 \label{A13} \\                                  &&\qquad \qquad \qquad \times 
    {q^2/4-({\bf p}+{\bf k})^2 \over i\omega/2\nu_0+q^2/4+({\bf p}+{\bf k})^2}
     {({\bf q}{\bf k})/2-({\bf p}{\bf k})-({\bf q}/2-{\bf p})^2 \over 
        i\omega/\nu_0+({\bf q}/2-{\bf p}-{\bf k})^2+({\bf q}/2+{\bf p})^2+k^2}
                                                             \, , \nonumber \\
   &&(i) = - {\lambda_0^4 D_0^2 \over 8 \nu_0^5} \int_p {q^2+2({\bf q}{\bf p}) 
                       \over ({\bf q}/2+{\bf p})^2 (i\omega/2\nu_0+q^2/4+p^2)} 
    \int_k {({\bf q}{\bf k})^2/4-({\bf p}{\bf k})^2 \over k^2} \times
 \label{A14} \\                                  &&\qquad \qquad \qquad \times 
    {q^2/4-({\bf p}+{\bf k})^2 \over i\omega/2\nu_0+q^2/4+({\bf p}+{\bf k})^2}
{1 \over i\omega/\nu_0+({\bf q}/2-{\bf p}-{\bf k})^2+({\bf q}/2+{\bf p})^2+k^2}
                                                               - \nonumber \\
  &&\quad - {\lambda_0^4 D_0^2 \over 8 \nu_0^5} \int_p 
{q^2-2({\bf q}{\bf p}) \over ({\bf q}/2-{\bf p})^2 (i\omega/2\nu_0+q^2/4+p^2)} 
  \int_k {[({\bf q}{\bf k})^2/4-({\bf p}{\bf k})^2][q^2/4-({\bf p}+{\bf k})^2]
  \over k^2 [i\omega/2\nu_0+q^2/4+({\bf p}+{\bf k})^2]} \times \nonumber \\ &&
\quad \times {1 \over ({\bf q}/2-{\bf p}-{\bf k})^2+({\bf q}/2-{\bf p})^2+k^2} 
   \biggl( 1 + {2 ({\bf q}/2-{\bf p})^2 \over  
 i\omega/\nu_0+({\bf q}/2-{\bf p}-{\bf k})^2+({\bf q}/2+{\bf p})^2+k^2} \biggr)
                                                             \, , \nonumber \\
   &&(k) = {\lambda_0^4 D_0^2 \over 4 \nu_0^5} 
    \int_p {q^2-2({\bf q}{\bf p}) \over ({\bf q}/2-{\bf p})^2} 
    \int_k {({\bf q}{\bf k})/2+({\bf p}{\bf k}) \over 
                                     k^2 ({\bf q}/2-{\bf p}-{\bf k})^2} 
    {q^2/4-({\bf p}+{\bf k})^2 \over i\omega/2\nu_0+q^2/4+({\bf p}+{\bf k})^2}
                                                                        \times 
 \label{A15} \\
  &&\qquad \times {({\bf q}{\bf k})/2-({\bf p}{\bf k})-k^2 \over 
                      ({\bf q}/2-{\bf p}-{\bf k})^2+({\bf q}/2-{\bf p})^2+k^2}
     {({\bf q}/2-{\bf p})^2 \over 
        i\omega/\nu_0+({\bf q}/2-{\bf p}-{\bf k})^2+({\bf q}/2+{\bf p})^2+k^2}
                                                             \, . \nonumber 
\end{eqnarray}

Upon collecting the several contributions, these expressions simplify 
considerably. One finds that the vertex functions may be split into 
UV--singular and UV--regular parts according to $\Gamma_{{\tilde h}{\tilde h}} 
= \Gamma_{{\tilde h}{\tilde h}}^{reg} + \Gamma_{{\tilde h}{\tilde h}}^{sing}$, 
and similarly $\Gamma_{{\tilde h} h} = \Gamma_{{\tilde h} h}^{reg} + 
\Gamma_{{\tilde h} h}^{sing}$. After some tedious, but elementary algebra one 
arrives at the following explicit results:
\begin{eqnarray}
   &&\Gamma_{{\tilde h}{\tilde h}}({\bf q},\omega)^{reg} = {\rm Re} 
    \biggl[ (i \omega + \nu_0 q^2) \biggl( {\lambda_0^4 D_0^3 \over 4 \nu_0^7} 
    \int_p {q^2/4-p^2 \over ({\bf q}/2+{\bf p})^2}  
                         {1 \over i\omega/2\nu_0+q^2/4+p^2} \times
 \label{A16} \\
   &&\qquad \qquad \qquad \times \int_k 
       {1 \over i\omega/2\nu_0+q^2/4+({\bf p}+{\bf k})^2}
 {({\bf q}{\bf k})/2+({\bf p}{\bf k}) \over k^2 ({\bf q}/2-{\bf p}-{\bf k})^2} 
                                                          \times \nonumber \\
   &&\qquad \qquad \qquad \qquad \qquad \qquad \qquad \qquad 
     \times {({\bf q}{\bf k})/2-({\bf p}{\bf k})-k^2 \over 
       i\omega/\nu_0+({\bf q}/2-{\bf p}-{\bf k})^2+({\bf q}/2+{\bf p})^2+k^2}
                                                                 \nonumber \\
   &&\quad + {\lambda_0^4 D_0^3 \over \nu_0^7} 
    \int_p {q^2/4-p^2 \over ({\bf q}/2+{\bf p})^2 ({\bf q}/2-{\bf p})^2} 
    \int_k {1 \over i\omega/2\nu_0+q^2/4+({\bf p}+{\bf k})^2} 
 {({\bf q}{\bf k})/2+({\bf p}{\bf k}) \over k^2 ({\bf q}/2-{\bf p}-{\bf k})^2}
 \times \nonumber \\                                          &&\qquad \times
 {({\bf q}/2-{\bf p})^2 k^2-[({\bf q}{\bf k})/2-({\bf p}{\bf k})]^2 \over
   ({\bf q}/2-{\bf p}-{\bf k})^2+({\bf q}/2-{\bf p})^2+k^2} \Bigl( {\rm Re} 
                         {1 \over i\omega/2\nu_0+q^2/4+p^2} +  \nonumber \\
   &&\qquad \qquad \qquad \qquad + 
     {1 \over i\omega/2\nu_0+q^2/4+p^2}{({\bf q}/2-{\bf p})^2 \over
         i\omega/\nu_0+({\bf q}/2-{\bf p}-{\bf k})^2+({\bf q}/2+{\bf p})^2+k^2}
                                   \Bigr) \biggr) \biggr] \, , \nonumber \\
   &&\Gamma_{{\tilde h}{\tilde h}}({\bf q},\omega)^{sing} = - 2 D_0 
    - {\lambda_0^2 D_0^2 \over 2 \nu_0^3} {\rm Re}  
  \int_p {(q^2/4-p^2)^2 \over ({\bf q}/2+{\bf p})^2 ({\bf q}/2-{\bf p})^2} 
         {1 \over i\omega/2\nu_0+q^2/4+p^2}
 \label{A17} \\
   &&\quad - {\lambda_0^4 D_0^3 \over 8 \nu_0^6} {\rm Re}
    \int_p {(q^2/4-p^2)^2 \over ({\bf q}/2+{\bf p})^2}
           {1 \over (i\omega/2\nu_0+q^2/4+p^2)^2}
    \int_k {({\bf q}/4-{\bf p}/2)^2-k^2 \over 
 ({\bf q}/4-{\bf p}/2+{\bf k})^2 ({\bf q}/4-{\bf p}/2-{\bf k})^2} \nonumber \\
   &&\quad - {\lambda_0^4 D_0^3 \over 4 \nu_0^6} {\rm Re}
    \int_p {(q^2/4-p^2)^2 \over ({\bf q}/2+{\bf p})^2 ({\bf q}/2-{\bf p})^2}
           {1 \over (i\omega/2\nu_0+q^2/4+p^2)^2} \times          \nonumber \\
   &&\qquad \qquad \qquad \qquad \qquad \qquad \qquad \qquad \times \int_k
    {({\bf q}/2-{\bf p})^2 k^2-[({\bf q}{\bf k})/2-({\bf p}{\bf k})]^2 \over 
 ({\bf q}/4-{\bf p}/2+{\bf k})^2 ({\bf q}/4-{\bf p}/2-{\bf k})^2} \nonumber \\
   &&\quad - {\lambda_0^4 D_0^3 \over 2 \nu_0^6} {\rm Re}
    \int_p {(q^2/4-p^2)^2 \over ({\bf q}/2+{\bf p})^2 ({\bf q}/2-{\bf p})^4}
           {1 \over i\omega/2\nu_0+q^2/4+p^2} \times              \nonumber \\
   &&\qquad \qquad \qquad \qquad \qquad \qquad \qquad \qquad \times \int_k
    {({\bf q}/2-{\bf p})^2 k^2-[({\bf q}{\bf k})/2-({\bf p}{\bf k})]^2 \over 
 ({\bf q}/4-{\bf p}/2+{\bf k})^2 ({\bf q}/4-{\bf p}/2-{\bf k})^2} 
                                                            \, , \nonumber \\
   &&\Gamma_{{\tilde h} h}({\bf q},\omega)^{reg} = - (i \omega + \nu_0 q^2) 
     \biggl[ {\lambda_0^4 D_0^2 \over 8 \nu_0^6} 
    \int_p {q^2(q^2/4+p^2)-2({\bf q}{\bf p})^2 \over ({\bf q}/2+{\bf p})^2
                                       (i\omega/2\nu_0+q^2/4+p^2)} \times
 \label{A18} \\
   &&\qquad \qquad \qquad \times \int_k 
     {1 \over i\omega/2\nu_0+q^2/4+({\bf p}+{\bf k})^2}
 {({\bf q}{\bf k})/2+({\bf p}{\bf k}) \over k^2 ({\bf q}/2-{\bf p}-{\bf k})^2} 
                                                          \times \nonumber \\
   &&\qquad \qquad \qquad \qquad \qquad \qquad \qquad \qquad 
     \times {({\bf q}{\bf k})/2-({\bf p}{\bf k})-k^2 \over 
       i\omega/\nu_0+({\bf q}/2-{\bf p}-{\bf k})^2+({\bf q}/2+{\bf p})^2+k^2}
                                                                 \nonumber \\
   &&\quad + {\lambda_0^4 D_0^2 \over 4 \nu_0^6} 
    \int_p {q^2(q^2/4+p^2)-2({\bf q}{\bf p})^2 \over 
  ({\bf q}/2+{\bf p})^2 ({\bf q}/2-{\bf p})^2 (i\omega/2\nu_0+q^2/4+p^2)} 
 \int_k {1 \over i\omega/2\nu_0+q^2/4+({\bf p}+{\bf k})^2} \times \nonumber \\
 &&\qquad \qquad \qquad \times {({\bf q}{\bf k})/2+({\bf p}{\bf k}) \over 
                                            k^2 ({\bf q}/2-{\bf p}-{\bf k})^2}
 {({\bf q}/2-{\bf p})^2 k^2-[({\bf q}{\bf k})/2-({\bf p}{\bf k})]^2 \over
        i\omega/\nu_0+({\bf q}/2-{\bf p}-{\bf k})^2+({\bf q}/2+{\bf p})^2+k^2} 
                                                                 \nonumber \\
   &&\quad + {\lambda_0^4 D_0^2 \over 4 \nu_0^6} 
    \int_p {q^2-2({\bf q}{\bf p}) \over ({\bf q}/2-{\bf p})^2}
 \int_k {1 \over i\omega/2\nu_0+q^2/4+({\bf p}+{\bf k})^2} 
 {({\bf q}{\bf k})/2+({\bf p}{\bf k}) \over k^2 ({\bf q}/2-{\bf p}-{\bf k})^2}
                                                          \times \nonumber \\
   &&\qquad \times
 {({\bf q}/2-{\bf p})^2 k^2-[({\bf q}{\bf k})/2-({\bf p}{\bf k})]^2 \over
                    [({\bf q}/2-{\bf p}-{\bf k})^2+({\bf q}/2-{\bf p})^2+k^2]
      [i\omega/\nu_0+({\bf q}/2-{\bf p}-{\bf k})^2+({\bf q}/2+{\bf p})^2+k^2]}
                                                    \biggr] \, , \nonumber \\
   &&\Gamma_{{\tilde h} h}({\bf q},\omega)^{sing} = i \omega + \nu_0 q^2
    + {\lambda_0^2 D_0 \over 4 \nu_0^2} 
  \int_p {q^2/4-p^2 \over ({\bf q}/2+{\bf p})^2 ({\bf q}/2-{\bf p})^2} 
         {q^2(q^2/4+p^2)-2({\bf q}{\bf p})^2 \over i\omega/2\nu_0+q^2/4+p^2}
 \label{A19} \\
   &&\quad + {\lambda_0^4 D_0^2 \over 16 \nu_0^5}
    \int_p {q^2/4-p^2 \over ({\bf q}/2+{\bf p})^2}
      {q^2(q^2/4+p^2)-2({\bf q}{\bf p})^2 \over (i\omega/2\nu_0+q^2/4+p^2)^2}
    \int_k {({\bf q}/4-{\bf p}/2)^2-k^2 \over 
 ({\bf q}/4-{\bf p}/2+{\bf k})^2 ({\bf q}/4-{\bf p}/2-{\bf k})^2} \nonumber \\
   &&\quad + {\lambda_0^4 D_0^2 \over 8 \nu_0^5}
    \int_p {q^2/4-p^2 \over ({\bf q}/2+{\bf p})^2 ({\bf q}/2-{\bf p})^2}
      {q^2(q^2/4+p^2)-2({\bf q}{\bf p})^2 \over (i\omega/2\nu_0+q^2/4+p^2)^2} 
                                                           \times \nonumber \\
   &&\qquad \qquad \qquad \qquad \qquad \qquad \qquad \qquad \times \int_k
   {({\bf q}/2-{\bf p})^2 k^2-[({\bf q}{\bf k})/2-({\bf p}{\bf k})]^2 \over
 ({\bf q}/4-{\bf p}/2+{\bf k})^2 ({\bf q}/4-{\bf p}/2-{\bf k})^2} \nonumber \\
   &&\quad + {\lambda_0^4 D_0^2 \over 8 \nu_0^5}
    \int_p {q^2/4-p^2 \over ({\bf q}/2-{\bf p})^4}
           {q^2-2({\bf q}{\bf p}) \over i\omega/2\nu_0+q^2/4+p^2} \int_k
    {({\bf q}/2-{\bf p})^2 k^2-[({\bf q}{\bf k})/2-({\bf p}{\bf k})]^2 \over 
 ({\bf q}/4-{\bf p}/2+{\bf k})^2 ({\bf q}/4-{\bf p}/2-{\bf k})^2} \nonumber \\
   &&\qquad - {\lambda_0^4 D_0^2 \over 16 \nu_0^5}
    \int_p {q^2/4-p^2 \over ({\bf q}/2-{\bf p})^2}
   {q^2-2({\bf q}{\bf p}) \over i\omega/2\nu_0+q^2/4+p^2} \times \nonumber \\
   &&\qquad \qquad \qquad \qquad \qquad \times \int_k
    {({\bf q}/2-{\bf p})^2 k^2-[({\bf q}{\bf k})/2-({\bf p}{\bf k})]^2 \over 
 ({\bf q}/4-{\bf p}/2+{\bf k})^2 ({\bf q}/4-{\bf p}/2-{\bf k})^2
        [3({\bf q}/2-{\bf p})^2/4+k^2]}                      \, . \nonumber
\end{eqnarray}

A powerful check to these lengthy calculations is the investigation of the 
situation at $d=1$, where the above expressions simplify considerably. Using 
some elementary algebra again, one finds at $d=1$
\begin{equation}
   {1 \over \nu_0 q^2} {\rm Re} \Gamma_{{\tilde h} h}(q,\omega) = 
        - {1 \over 2 D_0} \Gamma_{{\tilde h}{\tilde h}}(q,\omega) \, ,
 \label{fdt1}
\end{equation}
which ensures the validity of the fluctuation--dissipation theorem, for both
renormalized momentum-- and frequency--dependent quantities $\nu(q,\omega)$ and
$D(q,\omega)$, defined as the left-- and right--hand sides of Eq.~(\ref{fdt1}),
respectively, coincide. The explicit result reads:
\begin{eqnarray}
 \nu(q,\omega) &&= \nu_0 \biggl[ 1 + {\lambda_0^2 D_0 \over 4 \nu_0^3} {\rm Re}
      \int_p {1 \over i\omega/2\nu_0+q^2/4+p^2}             
 \label{d=1} \\
      &&\qquad \qquad \qquad + {\lambda_0^4 D_0^2 \over 16 \nu_0^6} {\rm Re}
      \int_p {(q/2-p)^2 \over (i\omega/2\nu_0+q^2/4+p^2)^2} 
      \int_k {1 \over k(q/2-p-k)}                                \nonumber \\
      &&- {\lambda_0^4 D_0^2 \over 8 \nu_0^6} {\rm Re} \biggl( 
      (i\omega/\nu_0q^2) \int_p {q/2-p \over i\omega/2\nu_0+q^2/4+p^2} 
      \int_k {1 \over i\omega/2\nu_0+q^2/4+(p+k)^2}              \nonumber \\ 
      &&\qquad \qquad \qquad \times {1 \over q/2-p-k} 
        {1 \over i\omega/\nu_0+(q/2-p-k)^2+(q/2+p)^2+k^2} \biggr) \biggr] \, , 
 \nonumber
\end{eqnarray}
which already is a useful and interesting result of the two--loop calculation 
on its own standing.

We now return to the general $d$--dimensional case. In evaluating the 
UV--singular contributions, one has to be careful to choose a normalization 
point (NP) where either $q$ or $\omega$ are finite, in order not to interfere 
with the IR singularities, which would show up as poles in $1/(d-2)$, too. A 
convenient choice is NP: ${\bf q} = {\bf 0}$, $i \omega / 2 \nu = \mu^2$; thus
we find
\begin{eqnarray}
   \Gamma_{{\tilde h}{\tilde h}}({\bf q},\omega)^{sing} \vert_{NP} = - 2 D_0
    \Biggl[ 1 &&+ g_0 \int_p {1 \over \mu^2 Z_\nu + p^2} 
              + g_0^2 \int_p {p^2 \over (\mu^2 + p^2)^2} 
          \int_k {p^2/4-k^2 \over ({\bf p}/2+{\bf k})^2 ({\bf p}/2-{\bf k})^2} 
 \nonumber \\
            &&+ 2 g_0^2 \int_p {1 \over (\mu^2 + p^2)^2} 
                        \int_k {p^2 k^2 - ({\bf p}{\bf k})^2 \over 
                        ({\bf p}/2+{\bf k})^2 ({\bf p}/2-{\bf k})^2}
 \nonumber \\
            &&+ 4 g_0^2 \int_p {1 \over p^2 (\mu^2 + p^2)} 
                        \int_k {p^2 k^2 - ({\bf p}{\bf k})^2 \over 
                        ({\bf p}/2+{\bf k})^2 ({\bf p}/2-{\bf k})^2} 
              + {\cal O}(g_0^3) \Biggr] \, ,
 \label{A20}
\end{eqnarray}
\begin{eqnarray}
   {\partial \over \partial q^2} 
     \Gamma_{{\tilde h} h}&&({\bf q},\omega)^{sing} \vert_{NP} = \nu_0 \Biggl[
                      1 - {d-2 \over d} g_0 \int_p {1 \over \mu^2 Z_\nu + p^2}
 \nonumber \\
             &&- {d-2 \over d} g_0^2 \int_p {p^2 \over (\mu^2 + p^2)^2} 
          \int_k {p^2/4-k^2 \over ({\bf p}/2+{\bf k})^2 ({\bf p}/2-{\bf k})^2} 
 \nonumber \\
             &&- 2 {d-2 \over d} g_0^2 \int_p {1 \over (\mu^2 + p^2)^2} 
                                  \int_k {p^2 k^2 - ({\bf p}{\bf k})^2 \over 
                                  ({\bf p}/2+{\bf k})^2 ({\bf p}/2-{\bf k})^2}
 \nonumber \\
             &&- 2 {d-2 \over d} g_0^2 \int_p {1 \over p^2 (\mu^2 + p^2)} 
                                  \int_k {p^2 k^2 - ({\bf p}{\bf k})^2 \over 
                                  ({\bf p}/2+{\bf k})^2 ({\bf p}/2-{\bf k})^2} 
 \nonumber \\
             &&+ g_0^2 \int_p {1 \over \mu^2 + p^2} 
                       \int_k {p^2 k^2 - ({\bf p}{\bf k})^2 \over 
                      ({\bf p}/2+{\bf k})^2 ({\bf p}/2-{\bf k})^2 (3p^2/4+k^2)}
 \nonumber \\
             &&+ {2 \over d} g_0^2 \int_p {1 \over \mu^2 + p^2} 
                              \int_k {[p^2 k^2 - ({\bf p}{\bf k})^2] k^2 \over 
                    ({\bf p}/2+{\bf k})^2 ({\bf p}/2-{\bf k})^2 (3p^2/4+k^2)^2}
 \nonumber \\
             &&+ {4 \over d} g_0^2 \int_p {1 \over p^2 (\mu^2 + p^2)} 
          \int_k {(p^2/4-k^2) [p^2 k^2 - ({\bf p}{\bf k})^2] ({\bf p}{\bf k})^2
                \over ({\bf p}/2+{\bf k})^4 ({\bf p}/2-{\bf k})^4 (3p^2/4+k^2)}
               + {\cal O}(g_0^3) \Biggr] \, .
 \label{A21}
\end{eqnarray}

For the derivation of the latter expression, the following relations have 
proven very useful:
\begin{eqnarray}
    \int_p ({\bf q}{\bf p}) f({\bf p}) &&= {q^2 \over d} \int_p p^2 f(p) \, ,
 \label{A22} \\
    \int_p \int_k ({\bf q}{\bf p}) ({\bf q}{\bf k}) ({\bf p}{\bf k}) 
                                                     f({\bf p},{\bf k}) &&= 
    {q^2 \over d} \int_p \int_k ({\bf p}{\bf k})^2 f({\bf p},{\bf k}) \, .
 \label{A23}
\end{eqnarray}

\subsection{Integrals in dimensional regularization}

In this appendix, we list the results for the above integrals, as obtained from
the dimensional regularization scheme \cite{dreg,amit}.

\begin{eqnarray}
   &&\int_p {1 \over \mu^2 + p^2} = - {C_d \mu^\epsilon \over \epsilon} \, ,
 \label{B1} \\
   &&\int_p {p^2 \over (\mu^2 + p^2)^2} 
     \int_k {p^2/4-k^2 \over ({\bf p}/2+{\bf k})^2 ({\bf p}/2-{\bf k})^2} 
     = - (d-1) {C_d^2 \mu^{2 \epsilon} \over 2 \epsilon^2}
      {\Gamma(1-\epsilon) \Gamma(1+\epsilon/2) \over \Gamma(1-\epsilon/2)} \, ,
 \label{B2} \\
   &&\int_p {1 \over (\mu^2 + p^2)^2} 
     \int_k {p^2 k^2 - ({\bf p}{\bf k})^2 \over 
                       ({\bf p}/2+{\bf k})^2 ({\bf p}/2-{\bf k})^2}     
     = (d-1) {C_d^2 \mu^{2 \epsilon} \over 4 \epsilon^2}
      {\Gamma(1-\epsilon) \Gamma(1+\epsilon/2) \over \Gamma(1-\epsilon/2)} \, ,
 \label{B3} \\
   &&\int_p {1 \over p^2 (\mu^2 + p^2)} 
     \int_k {p^2 k^2 - ({\bf p}{\bf k})^2 \over 
                       ({\bf p}/2+{\bf k})^2 ({\bf p}/2-{\bf k})^2} 
     = (d-1) {C_d^2 \mu^{2 \epsilon} \over 4 \epsilon^2 (1+\epsilon)}
      {\Gamma(1-\epsilon) \Gamma(1+\epsilon/2) \over \Gamma(1-\epsilon/2)} \, ,
 \label{B4} \\
   &&\int_p {1 \over \mu^2 + p^2} 
     \int_k {p^2 k^2 - ({\bf p}{\bf k})^2 \over 
                      ({\bf p}/2+{\bf k})^2 ({\bf p}/2-{\bf k})^2 (3p^2/4+k^2)}
 \nonumber \\
 &&\qquad \qquad \qquad \qquad \qquad \qquad \qquad \qquad = 
   - (d-1) {C_d^2 \mu^{2 \epsilon} \over 8 \epsilon} 
     {\Gamma(1-\epsilon) \Gamma(1+\epsilon) \over 
      \Gamma(1-\epsilon/2) \Gamma(1+\epsilon/2)} I_{00}(d) \, ,
 \label{B5} \\
   &&\int_p {1 \over \mu^2 + p^2} 
     \int_k {[p^2 k^2 - ({\bf p}{\bf k})^2] k^2 \over 
                    ({\bf p}/2+{\bf k})^2 ({\bf p}/2-{\bf k})^2 (3p^2/4+k^2)^2}
 \nonumber \\
   &&\quad = - (d-1) {C_d^2 \mu^{2 \epsilon} \over 16 \epsilon} 
    {\Gamma(1-\epsilon) \Gamma(1+\epsilon) \over 
     \Gamma(1-\epsilon/2) \Gamma(1+\epsilon/2)} 
     \left[ (d+\epsilon) I_{10}(d) + {2-\epsilon \over 4} I_{11}(d)
                + {2 - \epsilon \over 2} I_{21}(d) \right] \, ,
 \label{B6} \\
   &&\int_p {1 \over p^2 (\mu^2 + p^2)} 
    \int_k {(p^2/4-k^2) [p^2 k^2 - ({\bf p}{\bf k})^2] ({\bf p}{\bf k})^2 \over
                      ({\bf p}/2+{\bf k})^4 ({\bf p}/2-{\bf k})^4 (3p^2/4+k^2)}
 \nonumber \\
   &&\quad = (d-1) {(2-\epsilon) C_d^2 \mu^{2 \epsilon} \over 32
                      \epsilon} {\Gamma(1-\epsilon) \Gamma(1+\epsilon) \over 
                                 \Gamma(1-\epsilon/2) \Gamma(1+\epsilon/2)} 
    \Biggl[ (3-\epsilon) I_{00}(d) - {10-3\epsilon \over 2} I_{11}(d) 
 \nonumber \\     
 &&\qquad \; + {3-\epsilon \over 4} I_{21}(d) 
        + {7(4-\epsilon) \over 16} I_{22}(d) - {4-\epsilon \over 8} I_{32}(d) 
        + {1 \over 2} \int_0^1 {x(1-x) \over (1+x-x^2)^{2-\epsilon/2}} dx 
 \nonumber \\
&&\qquad \; - {4-\epsilon \over 4} [{\tilde I}_{01}(d) - {\tilde I}_{12}(d)]
 \Biggr] + (d-1) {C_d^2 \mu^{2 \epsilon} \over 16 \epsilon^2}
          {\Gamma(1-\epsilon) \Gamma(1+\epsilon/2) \over \Gamma(1-\epsilon/2)}
          \left( {2 \over 1+\epsilon} - (2 - \epsilon) \right) \, .
 \label{B7}
\end{eqnarray}

Here $C_d = \Gamma(2-d/2) / 2^{d-1} \pi^{d/2}$ is a geometry factor, while
\begin{equation}
   I_{rs}(d) = \int_0^1 \int_0^1 {y(1-y)^r \over 
                                  [x(1-x) y^2 + (1-y)(3+y)/4]^{s+2-d/2}} dx dy
 \label{B8}
\end{equation}
and
\begin{equation}
   {\tilde I}_{rs}(d) = 
    I_{rs}(d) - \int_0^1 \int_0^1 {(1-y)^r \over [x(1-x)+1-y]^{s+2-d/2}} dx dy
 \label{B9}
\end{equation}
are parameter integrals emerging upon the use of Feynman parametrization. 
Finally, we note that
\begin{eqnarray}
   {\Gamma(1-\epsilon) \Gamma(1+\epsilon/2) \over \Gamma(1-\epsilon/2)}
    &&= 1 + {\cal O}(\epsilon^2) \, ,
 \label{B10} \\
   {\Gamma(1-\epsilon) \Gamma(1+\epsilon) \over \Gamma(1-\epsilon/2)
    \Gamma(1+\epsilon/2)} &&= 1 + {\cal O}(\epsilon^2) \, ,
 \label{B11}
\end{eqnarray}
and
\begin{equation}
   \int_0^1 {x(1-x) \over (1+x-x^2)^2} dx = 
    {6 \over 5 \sqrt{5}} \ln {\sqrt{5} + 1 \over \sqrt{5} - 1} - {2 \over 5}
 \label{B12}
\end{equation}
are to be used when performing the $\epsilon$ expansion leading to the results 
for the Z factors. Note also that at $d=2$
\begin{eqnarray}
   &&8 I_{00}(2) - 4 I_{10}(2) - 21 I_{11}(2) + I_{21}(2) 
     + 7 I_{22}(2) - 2 I_{32}(2) \nonumber \\ 
  &&\quad - 4 {\tilde I}_{01}(2) + 4 {\tilde I}_{12}(2) + {12 \over 5 \sqrt{5}}
\ln {\sqrt{5} + 1 \over \sqrt{5} - 1} - {4 \over 5} + 4 = 8 - F_\nu(2) = 0 \, .
 \label{B13}
\end{eqnarray}

\vfill 

\eject

\newpage

\twocolumn

\narrowtext

\onecolumn
\newpage

\noindent {\bf Figure Captions:}

\vskip 1truecm

\begin{figure}
\figure{FIG.1: Basic elements of the dynamical perturbation theory for the 
	Burgers -- Kardar-Parisi-Zhang equation. (a) Correlation and response 
	propagators. (b) Three--point vertex.}
\label{basicelements}
\end{figure}

\begin{figure}
\figure{FIG.2: One--loop fixed point of the Burgers--KPZ equation as function 
	of the surface dimension $d$. The divergence at $d=3/2$ turns out to be
	an artefact of the one--loop approximation.}
\label{1-loop_FP}
\end{figure}

\begin{figure}
\figure{FIG.3: Two--loop fixed points of the Burgers--KPZ equation as functions
	of $d$ (full lines). It is not clear if the divergences at $d=1$ and 
	$d=2$ are inherent to the low order of perturbation expansion used, as
	was the case for the divergence of the one--loop fixed point (dashed 
	here) at $d=3/2$, or if they rather represent a generic feature of the 
        model.}
\label{2-loop_FP}
\end{figure}

\begin{figure}
\figure{FIG.4: The correlation length $\xi$ in the smooth phase versus 
        $|g-g_c^*|/g_c^*$. The correlation length crosses over from $\xi 
        \propto g^{1/\epsilon}$ at small couplings to $\xi \propto 
	|g-g_c^*|^{-\nu}$ as $g$ approaches the critical coupling $g_c^*$. In 
	the figure we have set $\epsilon = 1$.}
\label{correlation_length}
\end{figure}

\begin{figure}
\figure{FIG.5: The renormalized noise amplitude $D(g)$ (solid line) and surface
	tension $\nu(g)$ (dashed line) (up to a nonuniversal amplitude) in the 
	smooth phase versus $|g-g_c^*|/g_c^*$. The noise amplitude shows a 
	crossover from an exponential increase $D(g) \propto e^{g/\epsilon}$ at
	small values of $g$ to a power law divergence $D(g) \propto 
	|g-g_c^*|^{- \epsilon \nu}$ as $g$ approaches the critical coupling 
	$g_c^*$. The renormalized surface tension $\nu(g)$ crosses over from
        exponentially decreasing $\nu (g) \propto e^{-g/2}$ to a constant at
        criticality. The curves are plotted for $d=3$ (i.e. $\epsilon = 1$).}
\label{parameters}
\end{figure}

\begin{figure}
\figure{FIG.6: The characteristic frequency $\omega_c (g)$ (up to a 
	nonuniversal amplitude) in the smooth phase versus $|g-g_c^*|/g_c^*$. 
	The characteristic frequency crosses over from $\omega_c (g) \propto 
	g^{-2/\epsilon} e^{-g/2}$ at small couplings to $\omega_c (g) \propto 
	|g-g_c^*|^{-2}$ as $g$ approaches the critical coupling $g_c^*$. The 
	figure shows the result for $\epsilon = 1$.}
\label{char_freq}
\end{figure}

\begin{figure}
\figure{FIG.7: Feynman diagrams for the dynamical perturbation expansion of 
        $\Gamma_{{\tilde h}{\tilde h}}({\bf q},\omega)$ to two--loop order.}
\label{Gamma_02}
\end{figure}

\begin{figure}
\figure{FIG.8: Feynman diagrams for the dynamical perturbation expansion of 
        $\Gamma_{{\tilde h} h}({\bf q},\omega)$ to two--loop order.}
\label{Gamma_11}
\end{figure}


\begin{references}

\bibitem[1]{growth}
For a recent review, see: J. Krug and H. Spohn, in {\em Solids Far From 
Equilibrium: Growth, Morphology and Defects}, ed. C. Godriche (Cambridge 
University Press, Cambridge, 1992).

\bibitem[2]{surface}
See e.g., H. van Beijeren and I. Nolden, in {\em Structure and Dynamics of 
Surfaces II}, ed. by W. Schommers and P. von Blakenhagen (Springer-Verlag, 
Berlin, 1987);
see also T. Hwa, M. Kardar, and M. Paczuski, Phys. Rev. Lett. {\bf 66}, 441 
(1991), for a discussion of the nonequilibrium aspect.

\bibitem[3]{kpz86}
M. Kardar, G. Parisi, and Y.-C. Zhang, Phys. Rev. Lett. {\bf 56}, 889 (1986); 
E. Medina, T. Hwa, M. Kardar, and Y.-C. Zhang, Phys. Rev. {\bf A 39}, 3053 
(1989).

\bibitem[4]{fns77}
D. Forster, D. R. Nelson, and M. J. Stephen, Phys. Rev. A {\bf 16}, 732 (1977).

\bibitem[5]{dds}
H. van Beijeren, R. Kutner, and H. Spohn, Phys. Rev. Lett. {\bf 54}, 2026 
(1985);
H. K. Janssen and B. Schmittmann, Z. Phys. B {\bf 63}, 517 (1986).

\bibitem[6]{kus89}
J. Krug and H. Spohn, Europhys. Lett. {\bf 8}, 219 (1989).

\bibitem[7]{ks89}
V. S. L'vov and I. Procaccia, Phys. Rev. Lett. {\bf 69}, 3543 (1992);
I. Proccacia, M,H. Jensen, V.S. L'vov, K. Sneppen, and R. Zeitak, Phys. Rev.
A {\bf 46}, 3220 (1993).

\bibitem[8]{kz87}
M. Kardar and Y.-C. Zhang, Phys. Rev. Lett. {\bf 58}, 2087 (1987).

\bibitem[9]{sg89}
S. Zalesky, Physica D {\bf 34}, 417 (1989); and references therein.

\bibitem[10]{terry92}
T. Hwa, Phys. Rev. Lett. {\bf 69}, 1552 (1992).

\bibitem[11]{dp1}
D. A. Huse, C. L. Henley, and D. S. Fisher, Phys. Rev. Lett. {\bf 55}, 2924 
(1985); 
M. Kardar and Y.-C. Zhang, Phys. Rev. Lett. {\bf 58}, 2087 (1987).

\bibitem[12]{dp2}
For a recent review, see D. S. Fisher and D. A. Huse, Phys. Rev. B {\bf 43}, 
10728 (1991);
see also G. Parisi, J. Phys. (Paris) {\bf 51}, 1595 (1990); 
and M. Mezard, {\em ibid.}, 1831 (1990).

\bibitem[13]{ew}
S. F. Edwards and D. R. Wilkinson, Proc. R. Soc. London A {\bf 381}, 17 (1982).

\bibitem[14]{ft90}
B. M. Forrest and L.-H. Tang, Phys. Rev. Lett. {\bf 64}, 1405 (1990);
T. Ala--Nissila, T. Hjelt, and J. M. Kosterlitz, Europhys. Lett. {\bf 19}, 1 
(1992); 
T. Ala--Nissila, T. Hjelt, J. M. Kosterlitz, and O. Ven\"al\"ainen, J. Stat. 
Phys. {\bf 72}, 207 (1993).

\bibitem[15]{wk87}
D. E. Wolf and J. Kertesz, Europhys. Lett. {\bf 4}, 651 (1987).

\bibitem[16]{kk89}
J. M. Kim and J. M. Kosterlitz, Phys. Rev. Lett. {\bf 62}, 2289 (1989).

\bibitem[17]{fnote1}
Strictly speaking, the one--loop perturbation theory in 
Refs.~\cite{kpz86,fns77} gives a strong--coupling fixed point for $d < 3/2$ 
only. But, this is an artefact of the one--loop approximation and will be 
corrected by the two--loop calculations presented in this paper.

\bibitem[18]{dk92}
C. A. Doty and J. M. Kosterlitz, Phys. Rev. Lett. {\bf 69}, 1979 (1992).

\bibitem[19]{tnf90}
L.-H. Tang, T. Nattermann, and B. M. Forrest, Phys. Rev. Lett. {\bf 65}, 2422 
(1990).

\bibitem[20]{nat92}
T. Nattermann and L.-H. Tang, Phys. Rev. B {\bf 45}, 7156 (1992).

\bibitem[21]{j76}
H. K. Janssen, Z. Phys. B {\bf 23}, 377 (1976).

\bibitem[22]{bjw76}
R. Baussch, H. K. Janssen, and H. Wagner, Z. Phys. B {\bf 24}, 113 (1976).

\bibitem[23]{dreg}
G. t'Hooft and M. Veltman, Nucl. Phys. B {\bf 44}, 189 (1972).

\bibitem[24]{amit}
D. J. Amit, {\em Field Theory, the Renormalization Group, and Critical 
Phenomena}, 2nd ed. (World Scientific, Singapore, 1984).

\bibitem[25]{zinn_justin}
J. Zinn--Justin, {\em Quantum Field Theory and Critical Phenomena}, 
Clarendon Press, Oxford (1989).

\bibitem[26]{nsm}
A. M. Polyakov, Phys. Lett. 59 {\bf B}, 79 (1975);
E. Br\'ezin and J. Zinn--Justin, Phys. Rev. B {\bf 14}, 3110 (1976);
D. R. Nelson and R. A. Pelcovits, Phys. Rev. B {\bf 16}, 2191 (1977);
see also Ref.~\cite{amit}, Chap.~II.6, and Ref.~\cite{zinn_justin}, Chap.~27.

\bibitem[27]{sun94}
T. Sun and M. Plischke, {\em Field--Theory Renormalization Approach to the
Kardar--Parisi--Zhang Equation}, preprint (to be published in Phys. Rev. E, 
1994).

\bibitem[28]{dd75}
C. de Dominicis, Nuovo Cim. Lett. {\bf 12}, 567 (1975); 
J. Phys. (Paris) {\bf 37}, Colloque C--247 (1976).

\bibitem[29]{msr73}
P. C. Martin, E. D. Siggia, and H. H. Rose, Phys. Rev. A {\bf 8}, 423 (1973).

\bibitem[30]{dh75}
U. Deker and F. Haake, Phys. Rev. A {\bf 11}, 2043 (1975).

\bibitem[31]{g73}
R. Graham, in {\em Springer Tracts in Modern Physics}, Vol.~66, 
(Springer--Verlag, Berlin, 1973).

\bibitem[32]{mode_coupling}
For a review of the application of mode--coupling theories to dynamic critical 
phenomena see e.g., K. Kawasaki, in {\em Phase Transitions and Critical 
Phenomena}, Vol.~5a., eds. C. Domb and M.S. Green, Academic Press (1976).

\bibitem[33]{mode_kpz}
J. Krug, Phys. Rev. A {\bf 36}, 5465 (1987).

\bibitem[34]{erwin}
T. Hwa and E. Frey, Phys. Rev. A {\bf 44}, R7873 (1991).

\bibitem[35]{bc93}
J.P. Bouchaud and M.E. Cates, Phys. Rev. B {\bf 47}, R1455, (1993); Erratum 
Phys. Rev. B {\bf 48}, 635 (1993).

\bibitem[36]{dmkb94}
J.P. Doherty, M.A. Moore, J.M. Kim, and A.J. Bray, {\em Generalizations of the 
KPZ equation}, Phys. Rev. Lett. (in press).

\bibitem[37]{helge}
E. Frey and H. Schinz, (unpublished): We have repeated the numerical 
calculations of Refs.~\cite{bc93,dmkb94} and find agreement with the results of
Ref.~\cite{dmkb94}, but disagree with the upper critical dimension found in 
Ref.~\cite{bc93}.

\bibitem[38]{tu94}
Y. Tu, {\em Absence of finite upper critical dimension in the sperical KPZ 
model}, (unpublished, 1994).

\bibitem[39]{fnote2}
Sun and Plischke \cite{sun94}, however, do obviously not agree here. Instead, 
they claim a singular contribution to $\Gamma_{{\tilde h} h}({\bf 0},\omega)$.
To our opinion it is impossible to have such singular contributions for the 
following reasons. The identities in Section II.C are exact relations, which 
follow from Galilean invariance and the diffusive dynamics of the Burgers--KPZ 
equation. Those exact relations restrict the number of independent 
renormalization factors, as is well known from many other models in critical 
dynamics (see e.g. Ref.~\cite{bjw76}). In the dynamic functional for the 
Burgers--KPZ equation there are only four terms, which can be renormalized by 
the introduction of the {\it same} number of counter terms or (equally well) by
the {\it same} number of renormalization factors. Hence, if the theory is 
renormalizable, the maximum number of renormalization factors is four. We have 
introduced three, one for the noise amplitude, surface tension and
nonlinearity, respectively. In order to be complete, we could have introduced 
one more $Z$ factor to accout for possible renormalizations of the $i \omega$ 
term in the dynamic functional. The way to do that is to some extent arbitrary.
One choice would be to introduce an additional parameter in the $i \omega$ 
term, another choice is to introduce a renormalization factor for the fields 
$h$ and ${\tilde h}$. Whatever the choice would be, the exact relation in 
Section II.C, following from the diffusive behavior of the Burgers--KPZ 
equation, implies that the corresponding $Z$ factor equals $1$. Additionally, 
the Galilean invariance yields that the nonlinearity does not renormalize. 
Hence, we are left with {\it two} non--trivial renormalization factors.

\bibitem[40]{mama75}
S. K. Ma and G. F. Mazenko, Phys. Rev. B {\bf 11}, 4077 (1975).

\bibitem[41]{wiko74}
K. G. Wilson and J. Kogut, Phys. Rep. 12 {\bf C}, 75 (1974).

\bibitem[42]{itz_drouffe}
C. Itzykson and J.-M. Drouffe, ``{\em Statistical Field Theory}'', Vol.~1,
Cambridge University Press, Cambridge (1989).

\bibitem[43]{fnote3}
Sun and Plischke \cite{sun94} recover the FDT at $d=1$ by considering the 
one--dimensional case separately. Then they proceed with a $2-\epsilon$ 
expansion in the framework of which one would not expect the FDT to hold when 
one extrapolates to $\epsilon=1$. Note that we do not have to treat the 
one--dimensional case separately. Rather, it comes out quite naturally, if one 
makes a clear distinction between $1/\epsilon$ poles (UV singularities) and 
pure geometrical factors. The validity of the FDT is then a quite powerful 
check of the employed method.

\bibitem[44]{cayley1}
B. Derrida and H. Spohn, J. Stat. Phys.  {\bf 51}, 817 (1988).

\bibitem[45]{cayley2}
J. Cook and B. Derrida, Europhys. Lett. {\bf 10}, 195 (1989);
J. Phys. A {\bf 23}, 1523 (1990).

\bibitem[46]{is88}
J.Z. Imbrie and T. Spencer, J. Stat. Phys. {\bf 52}, 609 (1988).

\bibitem[47]{fht94}
E. Frey, T. Hwa, and U.C. T\"auber, (in preparation).

\bibitem[48]{fnote4}
There are several diagrams shown in the work by Sun and Plischke~\cite{sun94}, 
which violate causality, however, they are assigned zero value eventually, see
the appendix of Ref.~\cite{sun94}.

\end{references}
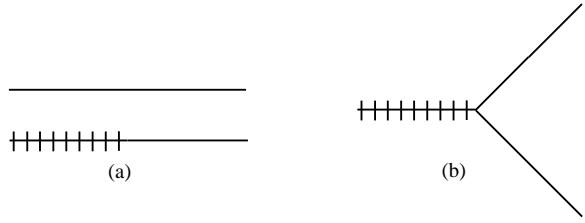
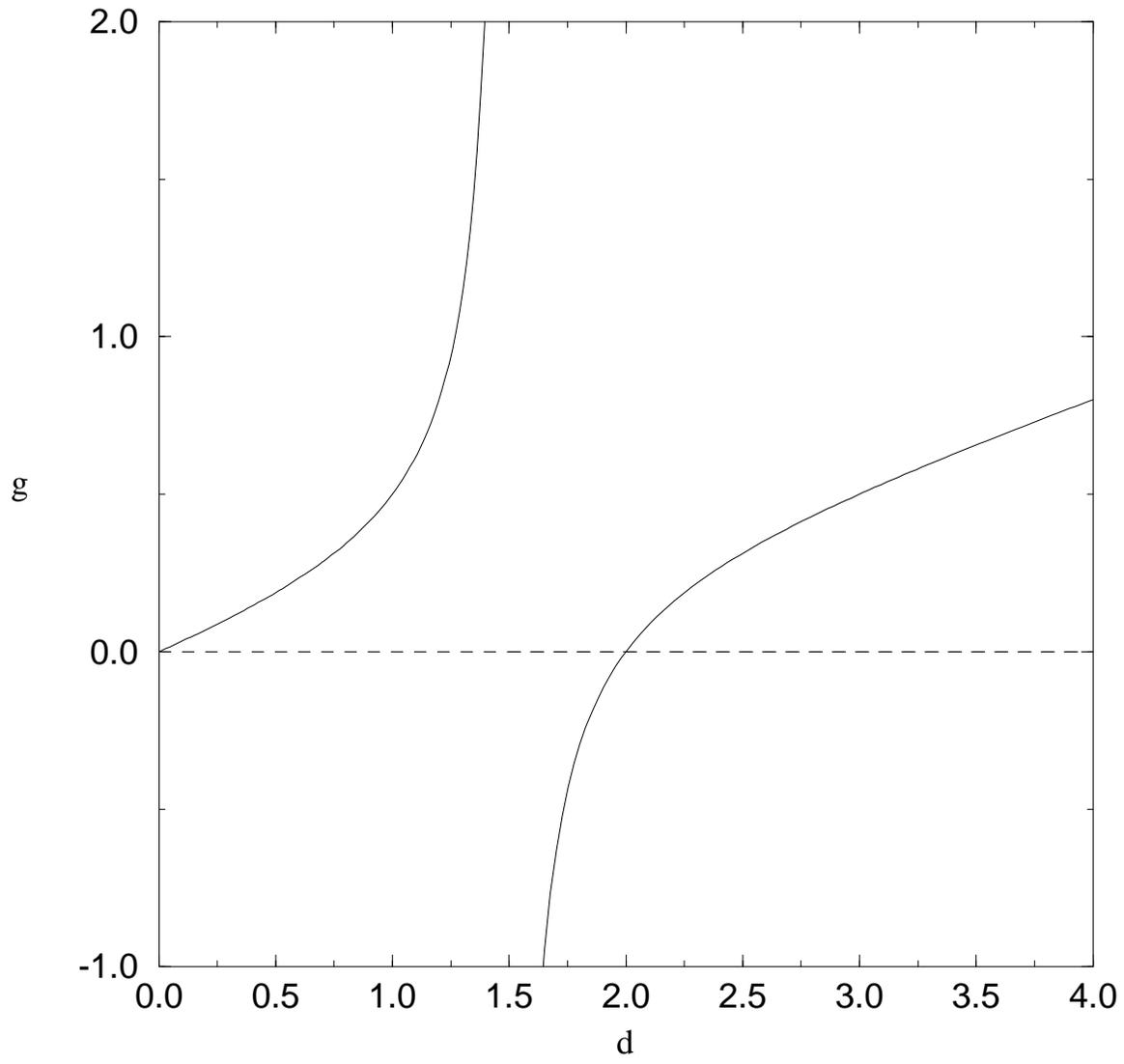
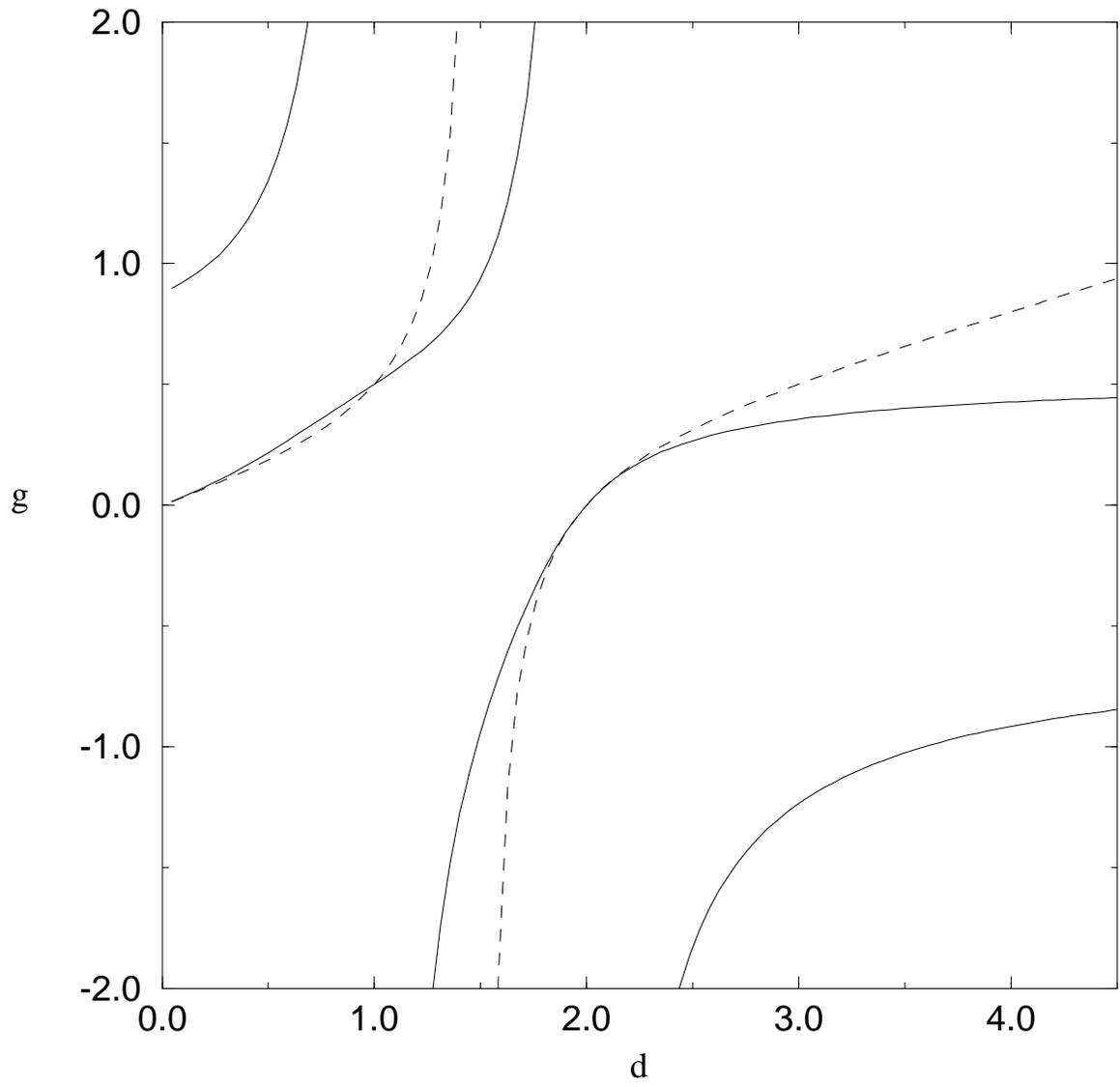
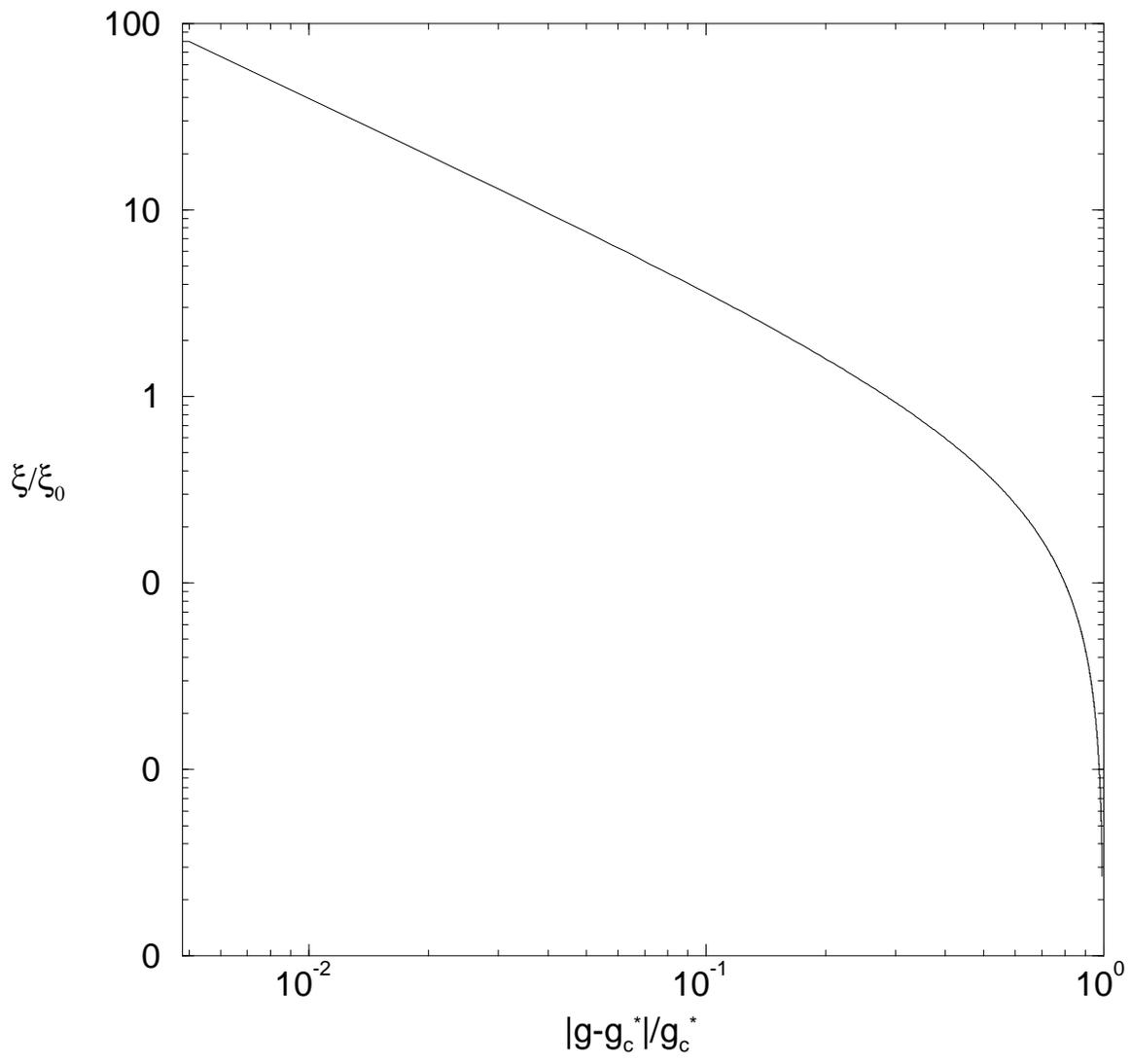
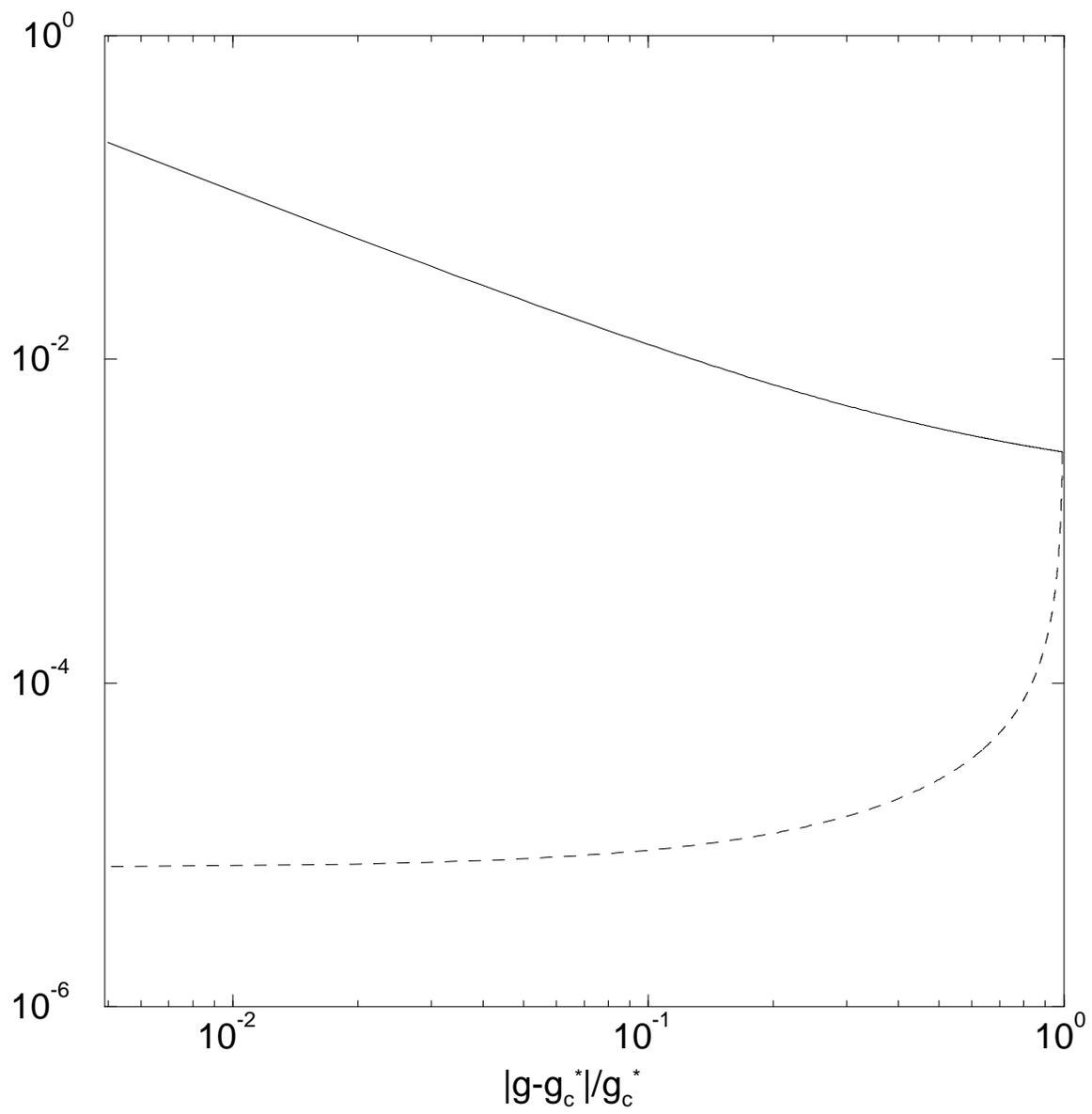
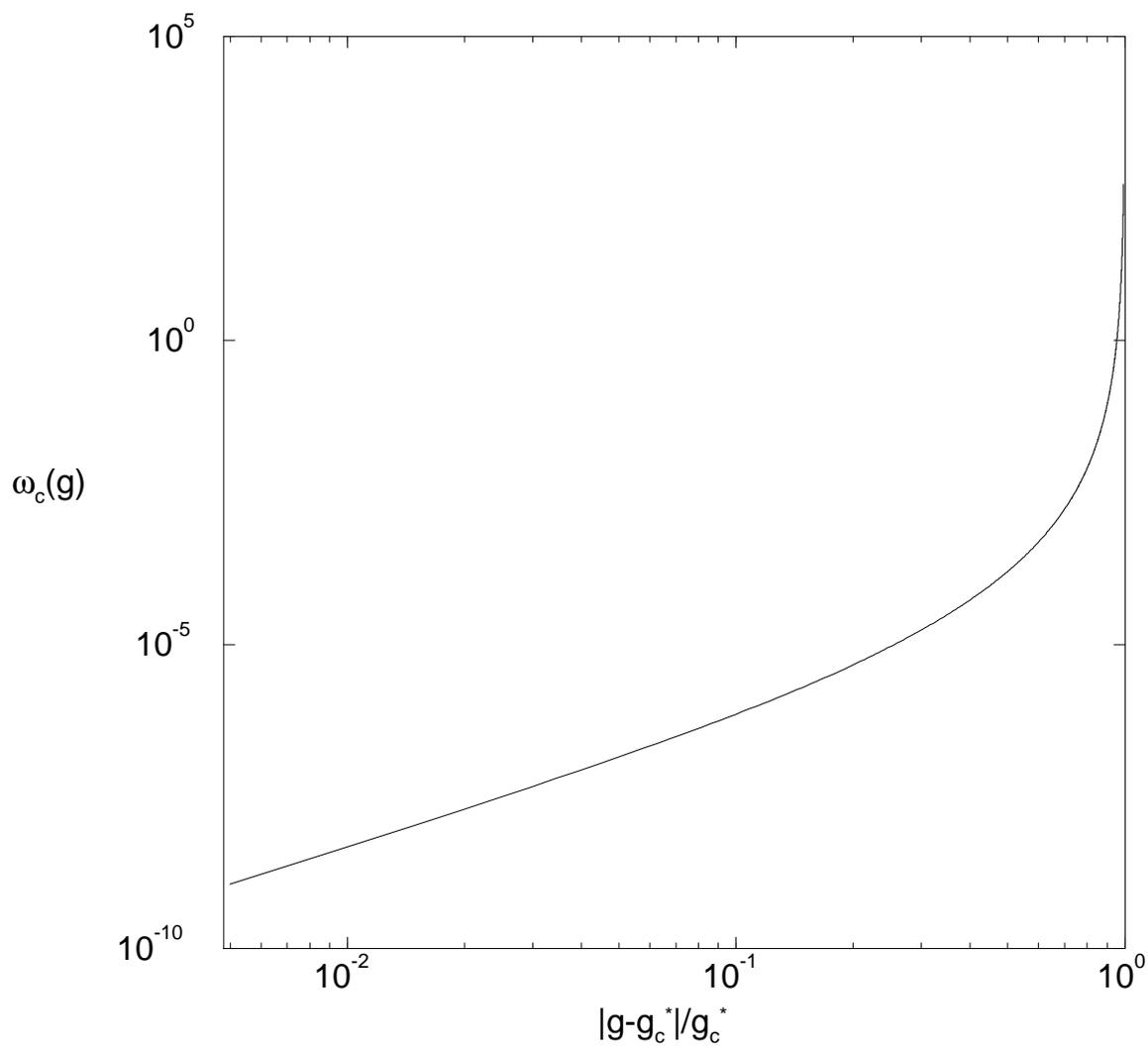
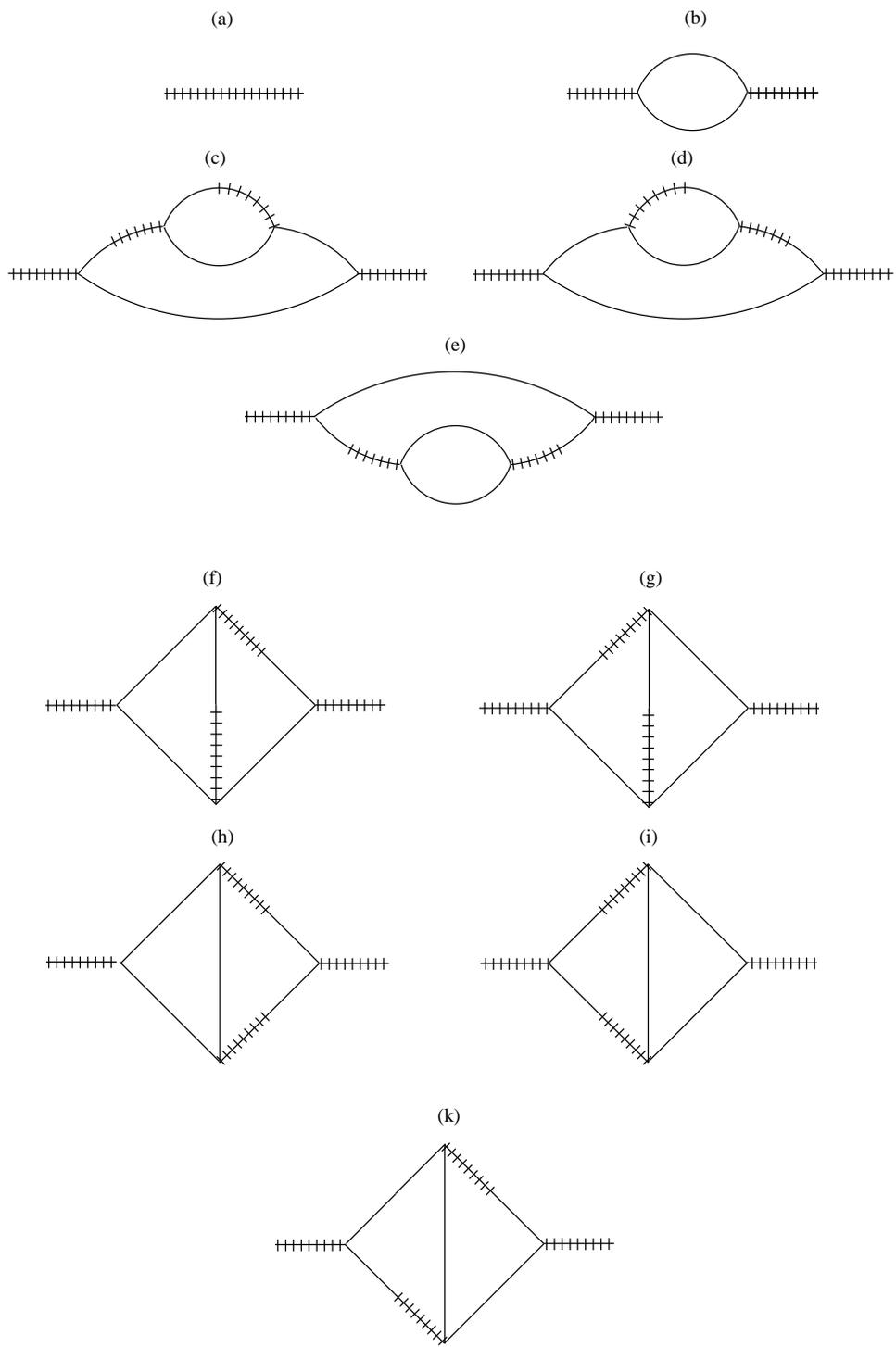
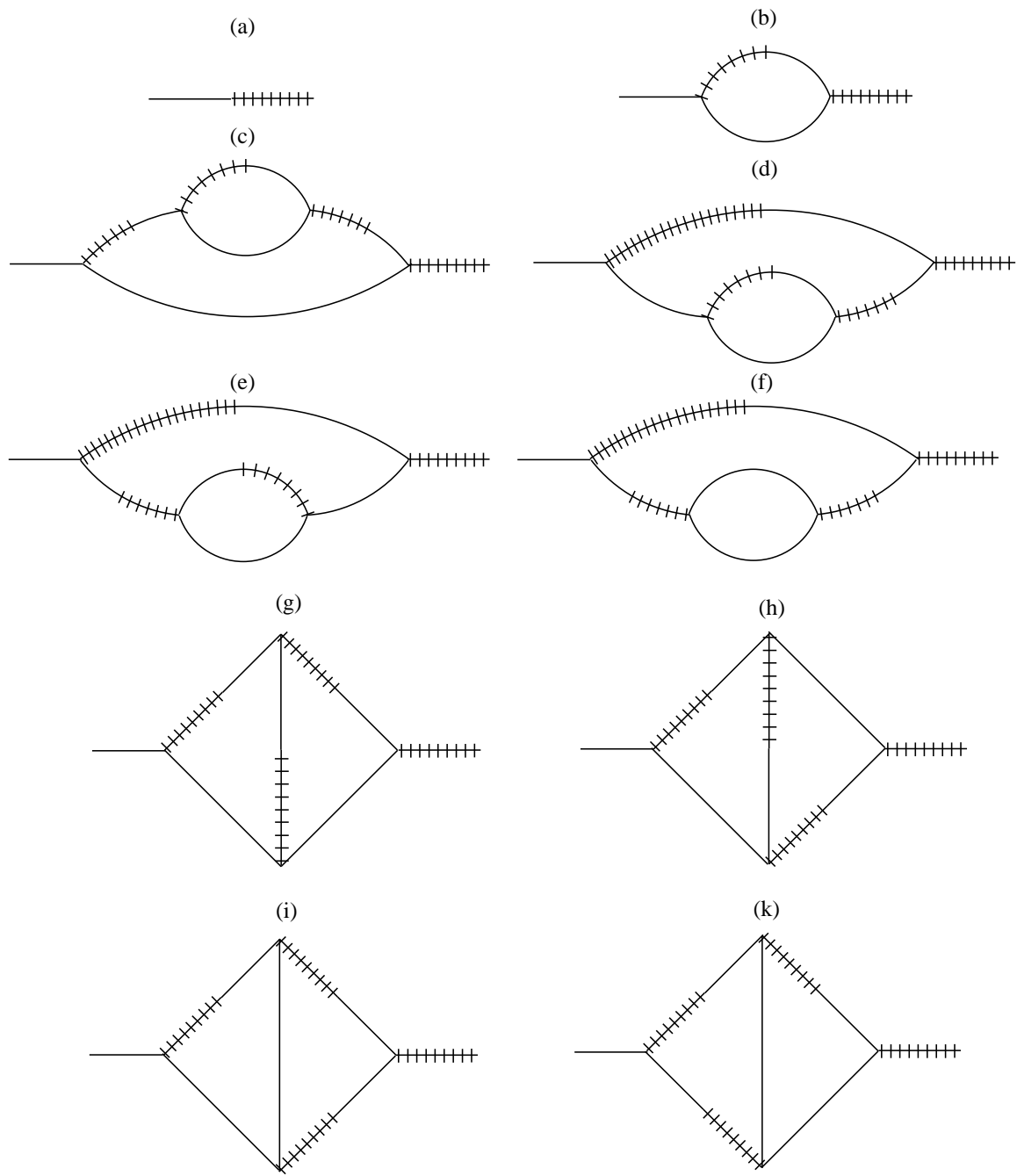
\end{document}